\documentclass[superscriptaddress,twocolumn,noshowpacs,longbibliography,amsmath,amssymb, amsfonts]{revtex4-1}   

\usepackage{graphicx}
\usepackage{dcolumn}
\usepackage{bm}
\usepackage{colordvi}
\usepackage{epsfig}
\usepackage[usenames]{color}
\usepackage{epstopdf}

\makeatletter
\renewcommand{\fnum@figure}{\textbf{Fig.~\thefigure}}
\renewcommand{\@caption@fignum@sep}{\ }
\makeatother

\makeatletter
\renewcommand{\fnum@table}{\textbf{Table~\thetable}}
\renewcommand{\@caption@fignum@sep}{\ }
\makeatother

\usepackage{titlesec}
\titleformat{\section}
  {\normalfont\normalsize\bfseries\raggedright}
  {\thesection}{1em}{}
\titlespacing*{\section}
  {0pt} 
  {4ex} 
  {0.3ex}  

\titleformat{\subsection}[runin]
  {\normalfont\normalsize\bfseries}
  {\thesubsection}
  {1em}
  {}
\titlespacing*{\subsection}
  {0pt}
  {4ex}
  {0.5em}
  
\usepackage{hyperref} 
\hypersetup{
  colorlinks,
  citecolor=blue,
  linkcolor=blue,
  urlcolor=blue
  }

\usepackage[usenames,dvipsnames]{xcolor}
\usepackage[normalem]{ulem}

\begin{document}
%
\title{Revealing the growth of structural length in glass-forming liquids without using dynamical information}
%
\author{
	Min~Liu$^{1,2\ast}$,
	Daisuke~Nishigaito$^{3}$,
    Yusuke~Hara$^{4}$,
    Norihiro~Oyama$^{4}$
    \& Takeshi~Kawasaki$^{1,2,3\dagger}$\\
	\small$^{1}$D3 Center, The University of Osaka, Toyonaka 560--0043, Japan.\\
	\small$^{2}$Department of Physics, The University of Osaka, Toyonaka 560--0043, Japan.\\
    \small$^{3}$Department of Physics, Nagoya University, Nagoya 464--8601, Japan.\\
    \small$^{4}$Toyota Central R\&D Labs., Inc., Nagakute 480--1192, Japan.\\
	\small$^\ast$email: liumin.d3c@osaka-u.ac.jp\\
    \small$^\dagger$email: kawasaki.takeshi.d3c@osaka-u.ac.jp\\
}

%
\begin{abstract}
Dynamical heterogeneity underlies the dramatic slowdown of supercooled liquids and has long been a central topic in glass physics. Despite considerable progress in identifying structural motifs and order parameters correlated with dynamical heterogeneity, a general characterization of structural descriptors remains elusive because they can vary among glass formers and are not generally known \emph{a priori}.
To address this challenge, we establish a unified framework that can be applied to varieties of glass formers to extract structural signatures of dynamical heterogeneity directly from their static configurations. The framework combines a graph neural network (GNN) with gradient-weighted class activation mapping (Grad-CAM).
Across three canonical three-dimensional glass-forming models, the resulting structural field exhibits a characteristic correlation length that tracks the corresponding dynamical correlation length.
Notably, a network trained on small systems can be applied to larger systems without retraining, indicating that the extracted features are encoded in local and mesoscopic structural correlations. 
These results demonstrate that static configurations encode structural signatures of dynamical heterogeneity and establish a scalable, interpretable route for extracting them.
\end{abstract}


\maketitle
%
\section*{Introduction}~\label{sec-introduction}
Upon slow cooling, a liquid undergoes a first-order phase transition into a crystalline solid, which is accompanied by the formation of periodic particle arrangements. In contrast, rapid cooling suppresses crystallization, producing a supercooled liquid below its melting temperature. In this regime, the viscosity increases sharply with decreasing temperature~\cite{Angell_1995g}, and the system eventually falls out of equilibrium to form a glass~\cite{Stillinger_1998g,Stillinger_2001g,Cavagna_2009g}. However, identifying clear microscopic structural differences between liquids and glasses is considerably more challenging than distinguishing liquids from crystalline solids~\cite{Kob_1995KAM,Dyre_2006g}. For example, standard structural quantities such as the static structure factor~\cite{Kob_1995sd,Kob_2002KAM} exhibit almost identical behavior in the liquid and glassy states. In contrast, their microscopic dynamics reveal a pronounced distinction: supercooled liquids exhibit spatially heterogeneous dynamics, characterized by the formation of mesoscale regions with high and low particle mobility. This phenomenon, known as dynamical heterogeneity~\cite{Hurley_1995dh,Kob_1997dh,Yamamoto_1998dh,Kegel_2000dh,Dauchot_2005dh}, becomes increasingly pronounced upon cooling, with a characteristic length scale that increases toward the glass transition~\cite{Lacevic_2003chi4}.
The extent to which dynamical heterogeneity can be inferred from static particle configurations remains a central question. Numerous studies have addressed this question by seeking structural descriptors that are associated with particle mobility. 

Early approaches largely relied on physically motivated order parameters, including locally favored structures, such as the icosahedral order in metallic glass formers and Lennard--Jones mixtures~\cite{Coslovich_2007,Cheng_2008,Malins_2013}, and medium-range crystalline order in polydisperse systems that are characterized by bond-orientational order parameters~\cite{Steinhardt_1983op1,Kawasaki_2007op2,Tanaka_2010op2,Kawasaki_2011op2,Tanaka_2012op1}. More recently, order parameters that do not presuppose specific structural motifs, such as a measure of the local steric order, have been demonstrated to correlate with the dynamics across several glass formers~\cite{Tong_2018TT,Tong_2019TT}. Although these studies have provided important insights, the choice of structural descriptors generally relies on prior physical considerations regarding the relevance of the structural features to particle dynamics. In addition, the extent to which a descriptor correlates with the dynamics may vary across glass-forming systems~\cite{Hocky_2014}.

\begin{figure*}
	\centering
	\includegraphics[width=1\textwidth]{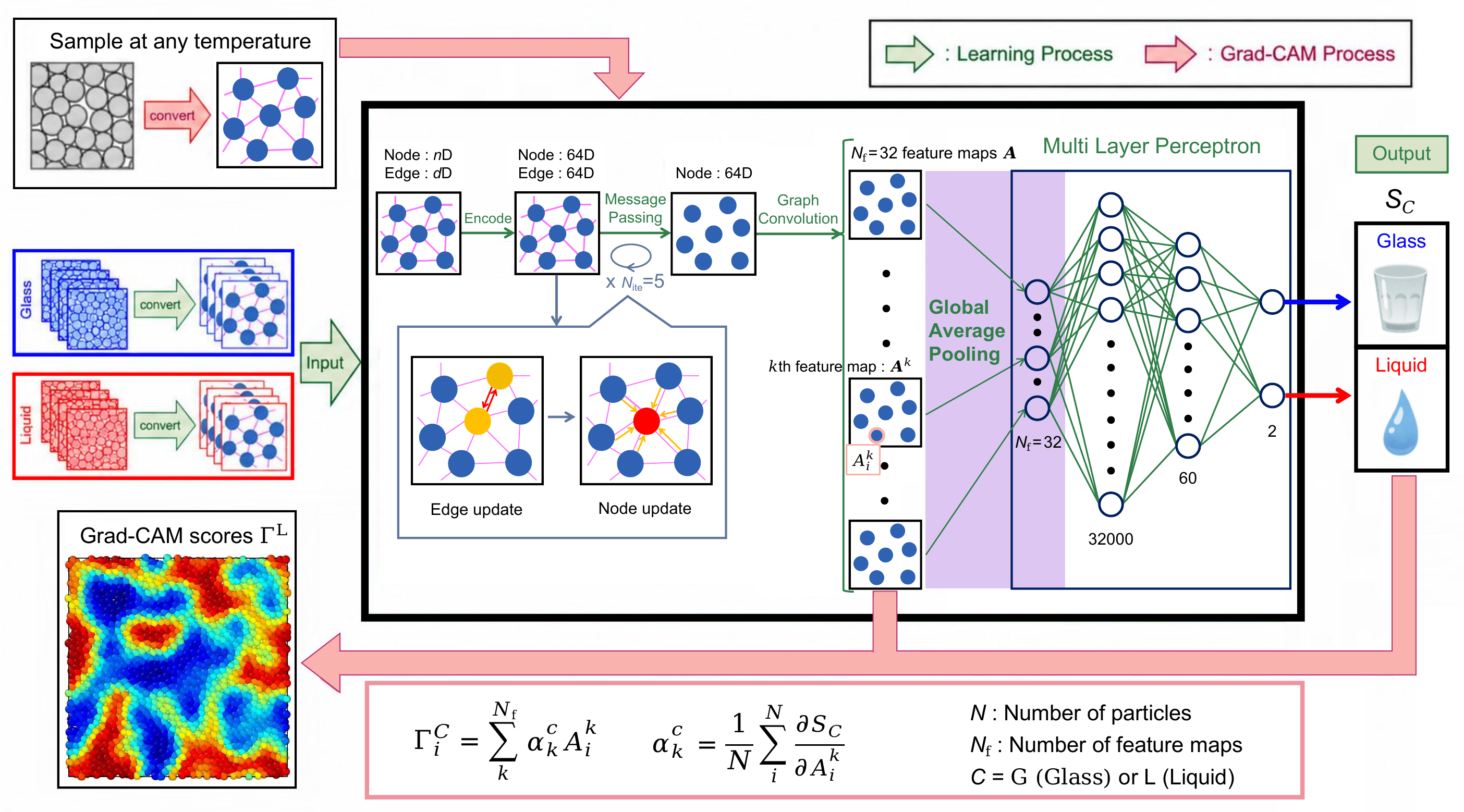} %
	\caption{\textbf{Schematic of the machine-learning method used for classification and structural interpretation.}
    The GNN is trained using labeled glass and liquid configurations; the green arrows indicate the learning process.
    Particle configurations are first converted into graph representations, in which the particles are treated as nodes and the neighboring particles are connected by edges.
    $n$D indicates the dimension of the node features~(the number of input particle features), whereas $d$D indicates the dimension of the edge features~(the dimension of the system).
    The initial node and edge features are encoded into $64$-dimensional~($64$D) embeddings, which are subsequently processed through $N_{\mathrm{ite}}=5$ message-passing iterations that comprise edge and node updates to obtain node representations with the same dimensions~($64$D).
    Thereafter, $N_\mathrm{f}=32$ feature maps $\boldsymbol{A}$ are generated by graph convolution operations.
    $A^k_i$ refers to the $i$th node in the $k$th feature map $\boldsymbol{A}^k$.
    The resulting feature maps are then pooled to $32$ values and fed to an MLP for classification.
    The values below each layer represent the number of outputs, and the final outputs are two classification scores $S_C$ ($C=$ G for glass or L for liquid), which successfully distinguish between the two states. 
    For the Grad-CAM analysis~(red arrows) of the particle configurations at arbitrary temperatures, we evaluated the gradients of the classification scores $S_C$ with respect to the feature maps $\boldsymbol{A}$. We then combined these to obtain the particle-based Grad-CAM scores $\Gamma^{C}_i$ (the formulas are shown at the bottom), producing a spatial map~(see $\Gamma^\mathrm{L}$ on the left as an example) that highlights the local regions that contribute most strongly to the classification.
    }
	\label{fig:gnn} %
\end{figure*}
\begin{figure*} 
	\centering
	\includegraphics[width=1\textwidth]{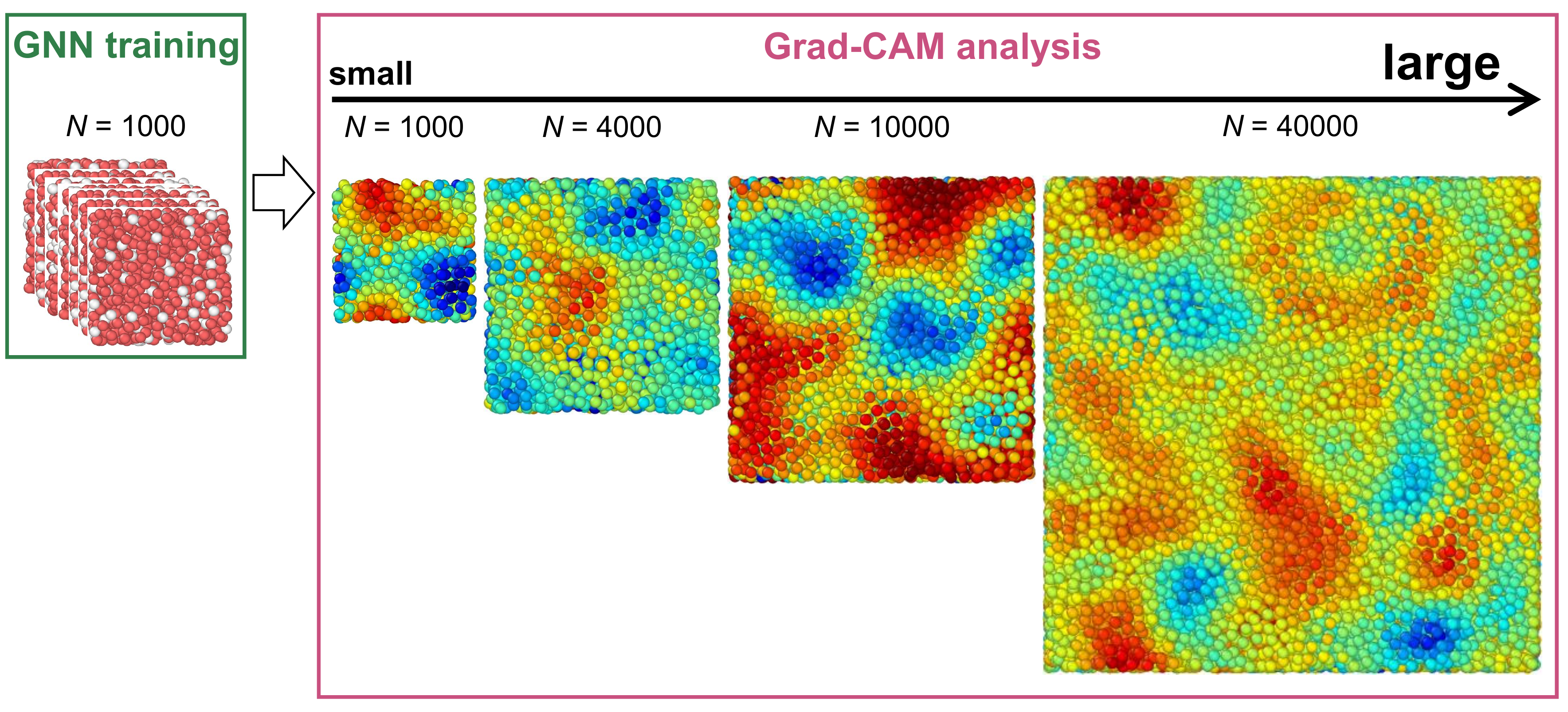} 
	\caption{\textbf{Training on small systems enables transferable structural analysis of substantially larger systems.}
	The GNN is trained only on configurations that contain $N=1000$ particles (left). The trained model and its Grad-CAM analysis are directly transferred to systems that contain up to $N=40000$ particles (red frame) without retraining or any change in the network parameters. The front views of the representative Grad-CAM structural fields $\Gamma^\mathrm{L}$ (see the text for the definition) for $N=1000$, $4000$, $10000$, and $40000$ are shown from left to right. This successful transfer across a broad range of system sizes shows that the learned structural field can be transferred without retraining and is consistent with a dominant contribution from transferable local and mesoscopic correlations, as opposed to system-size-specific global features. In addition, it substantially reduces the training cost while enabling the analysis of large-scale dynamical heterogeneity.}
	\label{fig:extend} %
\end{figure*}

Machine-learning approaches have been developed to learn the structural features that are relevant to dynamical heterogeneity and glass transition directly and reduce the dependence on such a priori selections~\cite{Cubuk_2015ML,Schoenholz_2016ML,Sussman_2017ML,Cubuk_2017ML,Schoenholz_2017ML,Ma_2019ML,Swanson_2020CNN,Landes_2020ML,Bapst_2020GNN,Paret_2020ML,Boattini_2020ML,Boattini_2021ML,Shiba_2023GNN,Kim_2023ML,Ciarella_2023ML,Jung_2023,Jung_2024ML,Jung_2025ML,Janzen_2024ML,Pezzicoli_2024GNN,Qiu_2025ML,Yoshikawa_2025GNN}. Many of these approaches train the models directly on dynamical observables such as the dynamic propensity. Neural networks that are trained in this manner, particularly graph neural networks~(GNNs), can identify the structural features that correlate strongly with the dynamics~\cite{Bapst_2020GNN,Shiba_2023GNN,Pezzicoli_2024GNN}.
Comparable predictive accuracy has also been achieved with simpler models based on locally averaged structural descriptors~\cite{Boattini_2021ML}.
Jung \textit{et al.} demonstrated that a propensity-trained model built on descriptors that are coarse grained over multiple lengths can be applied to substantially larger systems than those used for training, enabling the analysis of dynamical correlations beyond the training system size~\cite{Jung_2023,Jung_2024ML}. 
However, when dynamical observables are used as training labels, distinguishing the structural signatures associated with the dynamics from features that are selected through the supervision itself may be challenging.

Alternatively, relevant structural features can be sought from the static structure alone, without reference to dynamical observables. Boattini \textit{et al.} extracted the structural order parameters associated with dynamics in several three-dimensional (3D) glass formers via unsupervised learning of predefined bond-order descriptors~\cite{Boattini_2020ML}. Other unsupervised approaches have also been developed~\cite{Paret_2020ML}. Similarly, Qiu \textit{et al.} avoided propensity labels, deriving an order parameter that is correlated with the long-time dynamics obtained from short-time structural fluctuations and predefined radial and angular descriptors~\cite{Qiu_2025ML}. In these approaches, the input typically relies on predefined descriptors, and the relation between the learned representations and dynamical heterogeneity can be indirect. 

Training a classifier to distinguish liquid and glass configurations and then interpreting its decisions is a promising alternative. We previously combined a convolutional neural network (CNN) classifier for liquid and glass configurations~\cite{Swanson_2020CNN} with gradient-weighted class activation mapping (Grad-CAM)~\cite{Selvaraju_2020gradcam,Font-Clos_2022}. The findings showed that the resulting structural field correlates with dynamical heterogeneity following coarse graining~\cite{Oyama_2023,Liu_2024}.
However, this CNN-based approach requires the entire particle configuration to be converted into a grid-based density field and relies on a manually selected coarse-graining length. Beyond these limitations, to the best of our knowledge, no single framework has yet combined static-only training, input that is free of predefined descriptors or coarse-graining lengths, transferability across system sizes, and directly interpretable particle-resolved output. 

In this study, we demonstrate that these requirements can be met by combining a GNN with Grad-CAM.
The GNN directly takes particle species and coordinates as inputs, without requiring a grid-based representation. It is trained only to distinguish between glass and liquid configurations. Through message passing, the network progressively aggregates information from local particle neighborhoods. In this manner, the relevant spatial correlations are learned from the classification task, rather than prescribed through predefined descriptors or a fixed coarse-graining kernel. Subsequently, Grad-CAM is used to construct a particle-resolved structural field at arbitrary temperatures or packing fractions. The resulting field exhibits significant spatial correlations with dynamical heterogeneity on the structural-relaxation timescale as well as longer timescales. 
Importantly, we identify increasing static structural correlations that are strongly associated with dynamical heterogeneity in the super-Arrhenius regime across diverse glass-forming liquids. Although the local structural features that are selected by the network may differ among interaction models, their spatial organization provides a common structural signature that accompanies the emergence of slow heterogeneous dynamics.
Its characteristic correlation length increases upon cooling~(or compression) and follows the corresponding dynamical correlation length above the mode-coupling crossover in three glass-forming models. Furthermore, a network that is trained on relatively small systems can be applied directly to substantially larger systems without retraining, which is consistent with a dominant contribution from local and mesoscopic correlations. These results provide a scalable and interpretable framework for identifying the structural signatures that are associated with dynamical heterogeneity and for testing how mesoscopic structural correlations that are encoded in static configurations relate to slow dynamics.

\section*{Results}\label{section-result}
Many previous approaches to extracting the structural signatures of the long-time dynamics have relied on averaging or coarse graining the local structural descriptors over a prescribed length scale. Such averaging is often required to demonstrate the apparent correlations with the long-time dynamics, as in approaches based on bond-orientational order parameters, local softness-like indicators, and machine-learned structural representations~\cite{Tong_2018TT,Tong_2019TT,Boattini_2020ML}. Similarly, our previous CNN-based Grad-CAM analysis of the two-dimensional~(2D) Kob--Andersen (KA) model showed that the correspondence between the structural field \(\Gamma_{\rm CNN}^\mathrm{L}\) and dynamic propensity became apparent only following coarse graining~\cite{Liu_2024} (see Supplementary Fig.~1c, d). However, the appropriate averaging length is not known \emph{a priori} and may depend on the glass former and state point. 

In this study, we combined a GNN with Grad-CAM to construct particle-resolved structural fields from static configurations (Fig.~\ref{fig:gnn}; see also Methods). We applied this framework to three canonical glass-forming models: the 3D KA binary mixture, Wahnstr\"om~(WAH) binary mixture, and hard-sphere (HS)-like binary system with the Weeks--Chandler--Andersen repulsive potential~(WCA-HS). We extracted the Grad-CAM scores \(\Gamma^\mathrm{L}\) of the liquid-like structures for each model and examined their spatial correlations with the dynamical observables in supercooled liquids. 
The results for the 2D KA mixture showed spatial variations that are aligned with dynamic propensity without imposing a coarse-graining procedure, in contrast to the CNN-based method (Supplementary Fig.~1).

The Grad-CAM score $\Gamma^\mathrm{L}_i$ quantifies the contribution of the local environment of particle $i$ to the decision of the network that a configuration is liquid-like [equation~(\ref{eq:gradcam}) in Methods]. Because the network is trained only on equilibrium liquids at $T_\mathrm{L}$~(or $\eta^*_\mathrm{L}$) and on glasses obtained by cooling to $T_\mathrm{G}$~(or compressed to $\eta^*_\mathrm{G}$), a large (small) $\Gamma^\mathrm{L}_i$ indicates a local environment that is associated with the liquid (glass) reference state by the network. Thus, $\Gamma^\mathrm{L}$ locates each local environment between these two reference states for supercooled configurations at intermediate temperatures~(or packing fractions), to which neither label applies. We refer to particles with large and small $\Gamma^\mathrm{L}$ values as liquid-like and glass-like, respectively. This assignment is purely structural; its relation to mobility is examined in the following.

A key advantage of the present framework is its transferability across system sizes. The parameters of the GNN are independent of the number of nodes (or particles) owing to the global average pooling (highlighted in purple in Fig.~\ref{fig:gnn}). Thus, the GNN, which is trained only on configurations that contain \(N=1000\) particles, can be applied directly, without retraining, to substantially larger systems, including the largest configuration with \(N=40000\) particles studied in this case (Fig.~\ref{fig:extend}). 
This transferability substantially reduces the computational cost of training and is consistent with the learned structural field being dominated by transferable local and mesoscopic correlations, rather than by system-size-specific global features.
Next, we examine whether the transferred structural field reveals a mesoscopic organization that is associated with dynamical heterogeneity in larger systems. Unless stated otherwise, the following 3D results were obtained from samples with \(N=10000\), which is 10 times the particle number of the systems that were used for training.

\subsection*{Predicting dynamics of supercooled liquids.}
\begin{figure*} 
	\centering
	\includegraphics[width=1\textwidth]{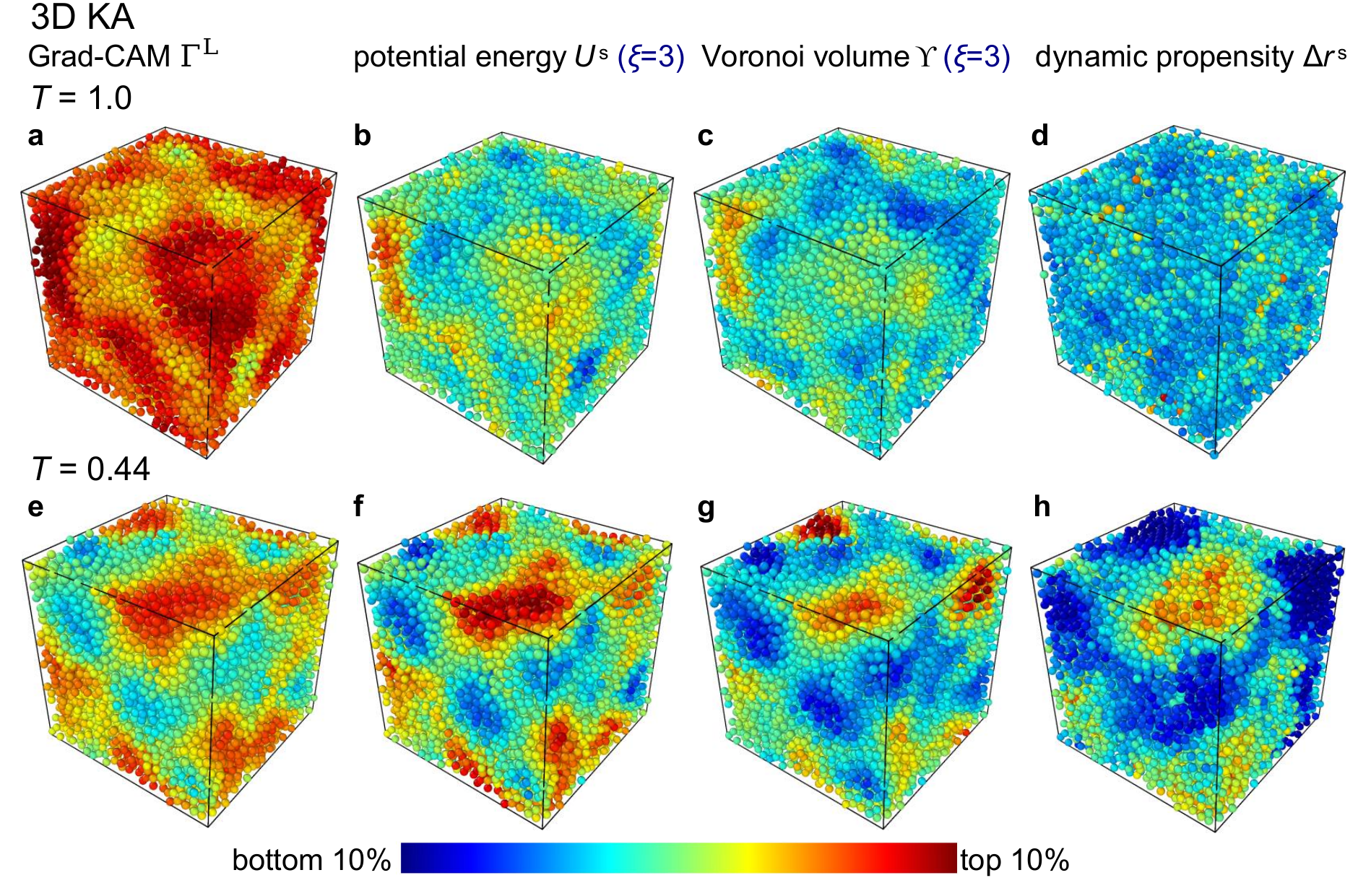} 
	\caption{\textbf{Snapshots of static and dynamical quantities in the 3D KA model.}
	From left to right: snapshots of the \textbf{a} Grad-CAM scores $\Gamma^\mathrm{L}$, \textbf{b} scaled potential energy $U^\mathrm{s}$, \textbf{c} Voronoi volume $\Upsilon$, and \textbf{d} scaled dynamic propensity $\Delta r^\mathrm{s}$~(which is explained in Methods and the Supplementary Materials) at $T = 1.0$ in the 3D KA system. \textbf{e}-\textbf{h} The same snapshots as in \textbf{a}-\textbf{d}, but at $T=0.44$. 
    All quantities shown are particle-resolved quantities.
    As the average value of the potential energy $U$ differed largely across temperatures, we show the scaled potential energy in (panels \textbf{b} and \textbf{f}), which was defined as $U^{\mathrm{s}}_i=U_i/\langle U \rangle$ for particle $i$, where $\langle U \rangle=N^{-1}\sum_{i=1}^{N} U_i$ is the mean potential energy of the configuration.
    The particle-resolved potential energy $U_i$ and Voronoi volume $\Upsilon_i$ (panels \textbf{c} and \textbf{g}) were coarse-grained using a coarse-graining length $\xi=3$. 
    Higher values of $\Gamma^\mathrm{L}$ indicate more liquid-like structures, whereas lower values indicate more glass-like structures. 
    Higher values of $\Delta r^\mathrm{s}$ correspond to more mobile particles, whereas lower values correspond to more immobile particles.
    The values increase with the color changes from blue to red, as indicated by the color bar at the bottom. 
    The leftmost blue corresponds to the global bottom 10\% of values across all samples at temperatures ranging from \(T_\mathrm{L}=1.0\) to \(T_\mathrm{G}=0.05\), whereas the rightmost red corresponds to the global top 10\%. 
    These configurations consisted of $N=10000$ particles, which were substantially larger than the samples used in the GNN training~(see Fig.~\ref{fig:extend}).}
	\label{fig:snap3d} %
\end{figure*}

\begin{figure*} 
	\centering
	\includegraphics[width=1\textwidth]{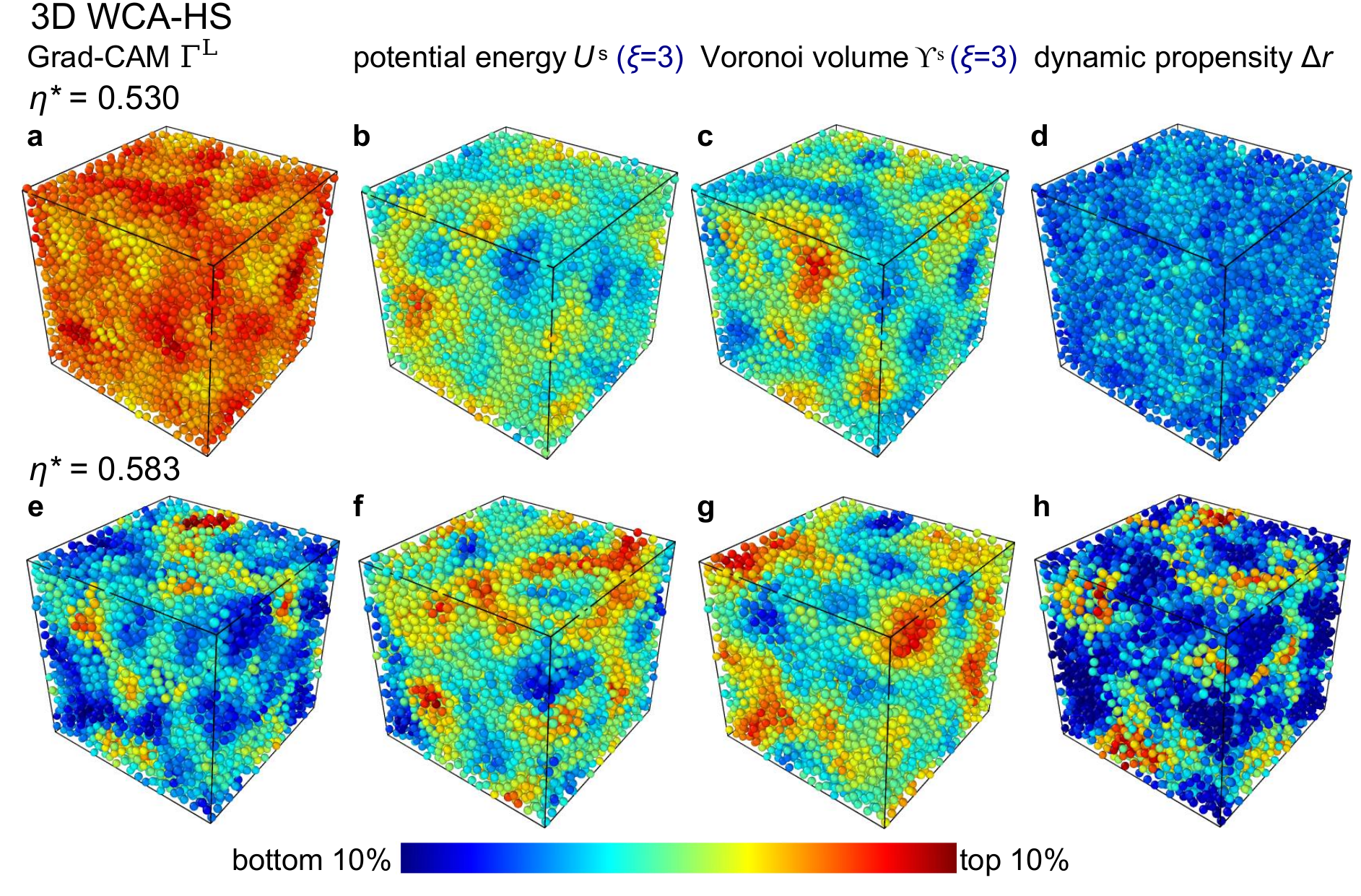} 
	\caption{\textbf{Snapshots of static and dynamical quantities in the 3D WCA-HS system.}
    From left to right: snapshots of the \textbf{a} Grad-CAM scores $\Gamma^\mathrm{L}$, \textbf{b} scaled potential energy $U^\mathrm{s}$, \textbf{c} scaled Voronoi volume $\Upsilon^\mathrm{s}$, and \textbf{d} dynamic propensity $\Delta r$ at $\eta^* = 0.530$ in the 3D WCA-HS model. \textbf{e}-\textbf{h} The same snapshots as in \textbf{a}-\textbf{d}, but with $\eta^* = 0.583$. 
    All quantities shown are particle-resolved quantities.
    In this case, the Voronoi volume $\Upsilon$ was also scaled owing to differences in the mean value at different packing fractions.
    The scaled potential energy (panels \textbf{b} and \textbf{f}) and scaled Voronoi volume (panels \textbf{c} and \textbf{g}) of particle $i$ can be obtained via $X^{\mathrm{s}}_i=X_i/\langle X \rangle$, where $X=U,\, \Upsilon$ and $\langle X \rangle=N^{-1}\sum_{i=1}^{N} X_i$ denotes the averaged potential or Voronoi volume of this configuration.
    The particle-resolved potential energy $U_i$ and Voronoi volume $\Upsilon_i$ were coarse-grained using a coarse-graining length $\xi=3$. 
    The values range from the global bottom 10\% value to the global top 10\% value, which were determined by all samples at packing fractions from $\eta^*_\mathrm{L}=0.493$ to $\eta^*_\mathrm{G}=0.70$.
    The colors of the values change from blue to red, as indicated by the color bar.
    These configurations consisted of $N=10000$ particles, larger than the samples for the GNN training.
	}
	\label{fig:snap3dhs} %
\end{figure*}

In this study, we first examined snapshots of the structural indicators and dynamic propensity in supercooled liquids. Both the Grad-CAM structural field and the dynamics for the 3D KA mixture appeared almost homogeneous at a high temperature of $T=1.0$. As shown in Fig.~\ref{fig:snap3d}a, the configuration was dominated by regions with a relatively high $\Gamma^\mathrm{L}$, which were classified as liquid-like by the GNN. The dynamic propensity $\Delta r^\mathrm{s}$ was also rather uniform spatially (Fig.~\ref{fig:snap3d}d). We analyzed the coarse-grained potential energy $U^\mathrm{s}$ (Fig.~\ref{fig:snap3d}b) and Voronoi volume $\Upsilon$ (Fig.~\ref{fig:snap3d}c), which have previously been reported to be associated with dynamical behavior~\cite{Berthier_2007iso,Liu_2024}, for comparison. These conventional indicators showed appreciable spatial variations even at this high temperature, without an equally apparent correspondence to the dynamic propensity.

The snapshots in Fig.~\ref{fig:snap3d}e-h show that pronounced spatial variations occurred in both the structural indicators and dynamics at a low temperature of $T=0.44$. Regions with a larger $\Gamma^\mathrm{L}$, which were classified as liquid-like by the GNN, were spatially separated from regions with a smaller $\Gamma^\mathrm{L}$, which were classified as glass-like (Fig.~\ref{fig:snap3d}e). In this configuration, the liquid-like regions also tended to exhibit a higher potential energy $U^\mathrm{s}$ (Fig.~\ref{fig:snap3d}f), larger Voronoi volume $\Upsilon$ (Fig.~\ref{fig:snap3d}g), and higher dynamic propensity (Fig.~\ref{fig:snap3d}h). Thus, the three structural indicators displayed qualitatively similar spatial patterns in the 3D KA mixture at a low temperature. The correspondence between the Grad-CAM field and dynamic propensity is quantified below.
The snapshots of the 3D WAH system are not presented here, as they exhibited behavior similar to that of the 3D KA mixture.

Next, we consider the 3D WCA-HS system. At a low packing fraction of $\eta^*=0.530$, the structural indicators and dynamic propensity (Fig.~\ref{fig:snap3dhs}a-d) showed distributions similar to those in the 3D KA model at a high temperature. At a higher packing fraction of $\eta^*=0.583$, the Grad-CAM scores $\Gamma^\mathrm{L}$ (Fig.~\ref{fig:snap3dhs}e) displayed a clear spatial correspondence with the dynamic propensity $\Delta r$ (Fig.~\ref{fig:snap3dhs}h). In contrast, the coarse-grained potential energy $U^\mathrm{s}$ (Fig.~\ref{fig:snap3dhs}f) exhibited no obvious spatial correspondence with either quantity. Instead, the coarse-grained Voronoi volume $\Upsilon^\mathrm{s}$ (Fig.~\ref{fig:snap3dhs}g) varied oppositely to $\Gamma^\mathrm{L}$ and $\Delta r$: regions with a lower $\Upsilon^\mathrm{s}$ tended to be liquid-like and more mobile, whereas regions with a higher $\Upsilon^\mathrm{s}$ tended to be glass-like and less mobile.

These comparisons illustrate that the spatial relationship between conventional structural indicators and dynamics is system-dependent. This is consistent with previous reports in which the correlation between the local order and mobility varied strongly across glass formers~\cite{Hocky_2014}. In particular, the apparent relation of the potential energy with the dynamics was weak in the 3D WCA-HS mixture, and the sign of the correspondence between the Voronoi volume and mobility differed between the 3D KA and 3D WCA-HS systems. In contrast, the Grad-CAM fields that were obtained using the same GNN architecture and training protocol showed a positive spatial correspondence with mobility in both systems.

\begin{figure} %
	\centering
	\includegraphics[width=0.43\textwidth]{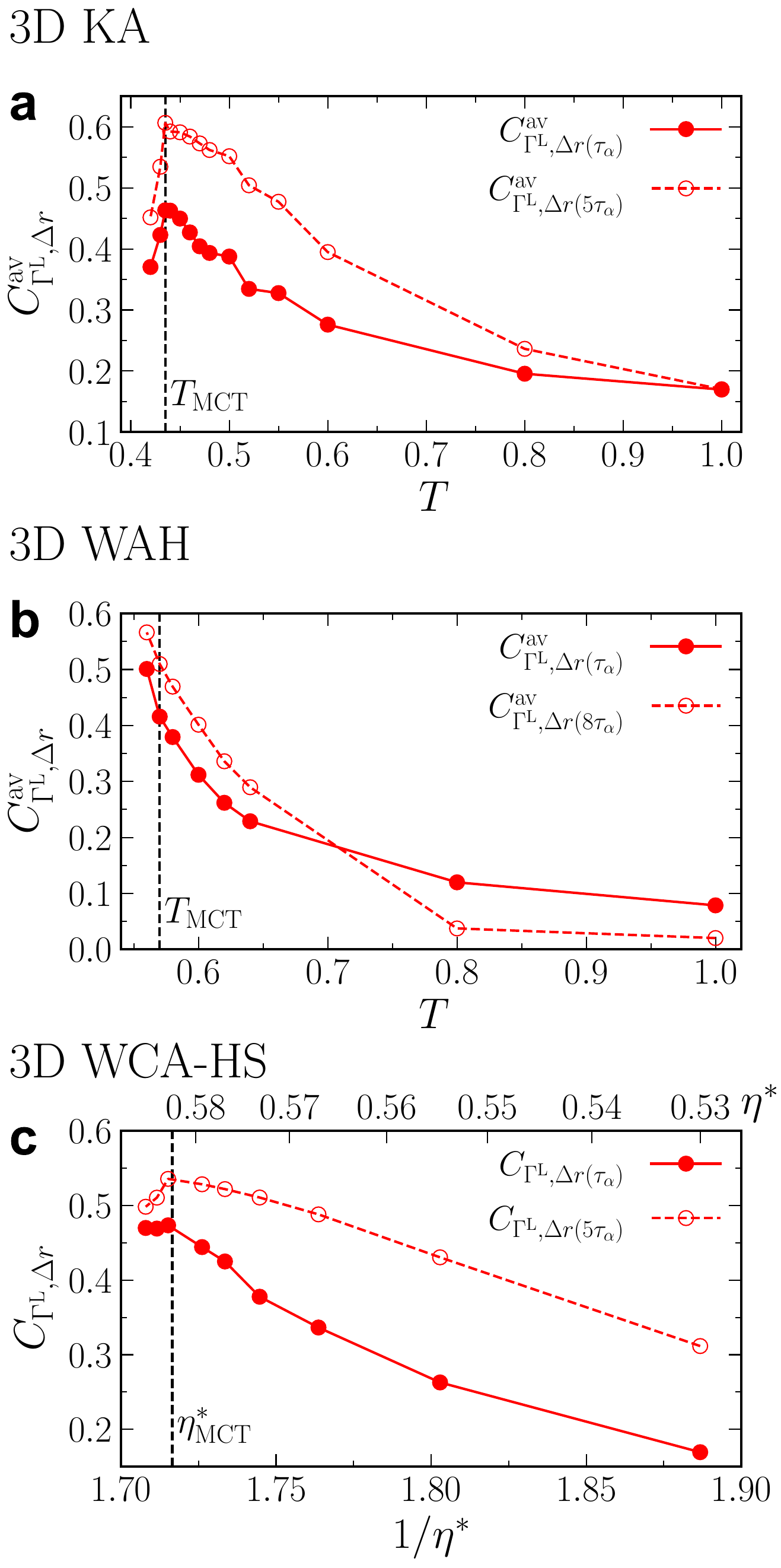}
	\caption{\textbf{Temperature/packing fraction dependence of structure--dynamics correlation coefficients.}
    Pearson's correlation coefficients $C^\mathrm{av}_{\Gamma^\mathrm{L},\Delta r}$ for the \textbf{a} 3D KA and \textbf{b} 3D WAH models, and $C_{\Gamma^\mathrm{L},\Delta r}$ for the \textbf{c} 3D WCA-HS model between the Grad-CAM scores $\Gamma^\mathrm{L}$ and dynamic propensity $\Delta r$ against temperatures $T$~(or the inverse effective packing fraction $1/\eta^*$ in \textbf{c}).
    The definitions of $C^\mathrm{av}_{\Gamma^\mathrm{L},\Delta r}$ and $C_{\Gamma^\mathrm{L},\Delta r}$ can be found in Methods.
    The solid circles denote correlations with the dynamic propensity $\Delta r$ over the timescale of $\tau_\alpha$, whereas the open circles denote those with $\Delta r$ over longer timescales ($5\tau_\alpha$ for \textbf{a}, \textbf{c}, and $8\tau_\alpha$ for \textbf{b}), where the correlations reached their maximum values (see Supplementary Fig.~11).
    In the 3D KA mixture (panel \textbf{a}), $T_{\mathrm{MCT}}=0.435$, indicated by the vertical dashed line, is the mode-coupling temperature.
    At $T_{\mathrm{MCT}}$, $C^{\mathrm{av}}_{\Gamma^{\mathrm{L}},\Delta r}$ attained its maximum value of 0.463, with $C^{\mathrm{A}}_{\Gamma^{\mathrm{L}},\Delta r}=0.448$ for larger A-particles and $C^{\mathrm{B}}_{\Gamma^{\mathrm{L}},\Delta r}=0.521$ for smaller B-particles.
    In the 3D WAH model (panel \textbf{b}), $T_{\mathrm{MCT}}=0.57$.
    In the 3D WCA-HS system (panel \textbf{c}), $\eta^*_{\mathrm{MCT}}=0.5826$.
    The values of the effective packing fraction $\eta^*$ are also shown at the top.}
	\label{fig:corref} 
\end{figure}

As shown in Fig.~\ref{fig:corref}, the Pearson's correlation coefficients generally increased upon cooling or compression in all three systems, indicating an increasingly strong spatial association between the Grad-CAM structural field and dynamical heterogeneity. The correlations reached their highest values near $T_{\mathrm{MCT}}$ or $\eta^*_{\mathrm{MCT}}$. 
The values decreased or saturated in the 3D KA and 3D WCA-HS models at lower temperatures or higher packing fractions, whereas the correlation in the 3D WAH system continued to increase over the investigated range. 
A direct comparison with previous machine-learning approaches that do not use dynamical information~\cite{Paret_2020ML,Boattini_2020ML} is complicated owing to differences in the correlation measures, particle species, timescales, coarse-graining procedures, and, in certain cases, the model itself.
For the 3D KA mixture at a comparable state point ($T = 0.5$), the present correlation with the dynamic propensity $\Delta r$ over the timescale of $\tau_\alpha$ was similar to that based on the unsupervised order parameter upon coarse graining~(the ``local average") in Ref.~\cite{Boattini_2020ML}, at approximately 0.4. 
However, studying $\Delta r$ over a longer timescale of $5\tau_\alpha$ revealed that it was enhanced significantly to 0.55, which is higher than the results of Ref.~\cite{Boattini_2020ML}.
The present framework yielded consistently positive correlations across all three models using a single architecture and learning protocol, without predefined descriptors or coarse graining.

\begin{figure*} 
	\centering
	\includegraphics[width=1\textwidth]{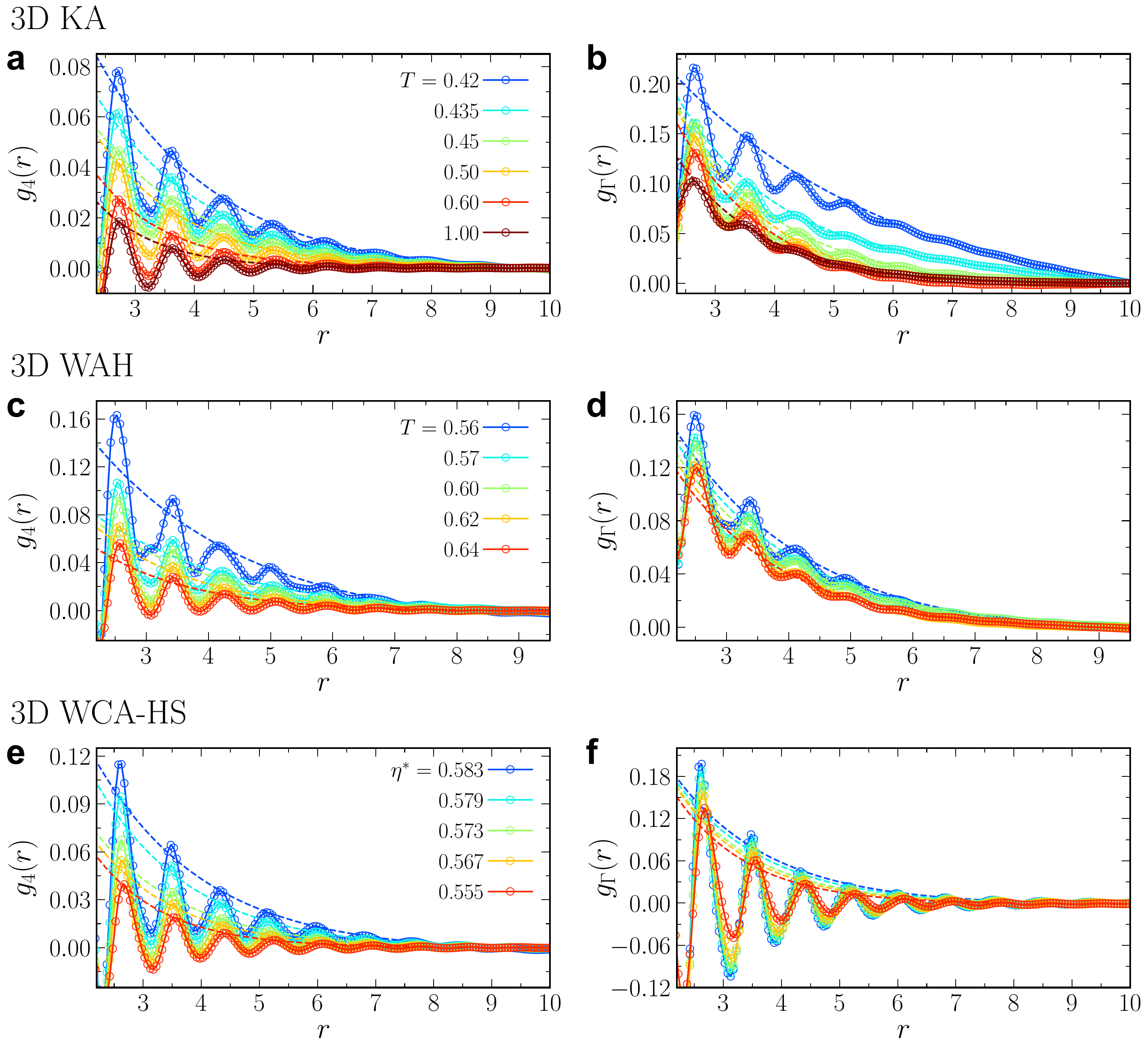}
	\caption{\textbf{Radial correlation functions of the dynamics and structures.}
	Radial correlation functions $g_{4}(r)$ of the slow dynamics at different temperatures or effective packing fractions $\eta^*$  (distinguished by colors) in the \textbf{a} 3D KA, \textbf{c} 3D WAH, and \textbf{e} 3D WCA-HS models.
    Radial correlation functions $g_{\Gamma}(r)$ of glass-like structures characterized by Grad-CAM scores at different temperatures~(or different packing fractions) in the \textbf{b} 3D KA, \textbf{d} 3D WAH, and \textbf{f} 3D WCA-HS models.
    The dashed curves correspond to the ``envelope fit" for the range of $r>5$ using functions proportional to $e^{(-r/\xi)}$, resulting in the correlation length $\xi$~(see Fig.~\ref{fig:correl}).
    }
	\label{fig:g4} %
\end{figure*}

\subsection*{Radial correlations of dynamics and structures.}

\begin{figure} %
	\centering
	\includegraphics[width=0.45\textwidth]{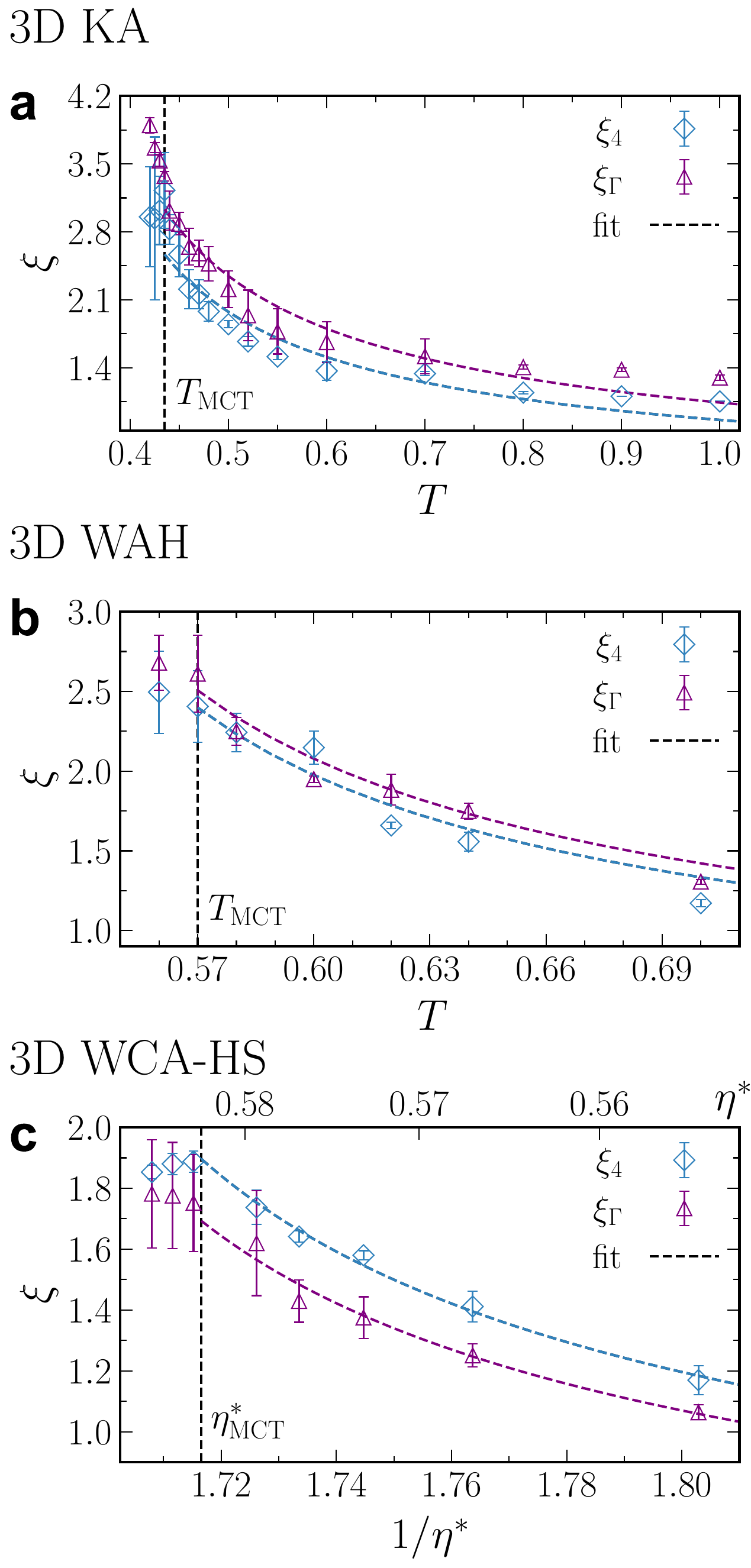}
	\caption{\textbf{Temperature/packing fraction dependence of the structural and dynamical correlation lengths.}
     \textbf{a} Correlation lengths of the dynamics $\xi_{4}$ (blue symbols) and those of the Grad-CAM scores $\xi_{\Gamma}$ (purple symbols) with error bars for the 3D KA mixture, obtained by fitting the data in Fig.~\ref{fig:g4}.
     The dashed curves denote fits to the data for $T>T_{\mathrm{MCT}}$ with $\xi=\xi_0[T_0/(T-T_0)]^\nu$, where $T_0=0.30$ is the extrapolation temperature obtained from VFT equation fits to the $T$ dependence of $\tau_\alpha$ (see Supplementary Fig.~4a). The fitted results provided $\nu=0.64$ for $\xi_\Gamma$ and $0.66$ for $\xi_4$. 
     \textbf{b} The same quantities as in (A) for the 3D WAH model. 
     Here, $T_0=0.49$, and the fitted $\nu=0.58$ for $\xi_\Gamma$ and $\nu=0.60$ for $\xi_4$.
     \textbf{c} Correlation length scales as a function of the effective packing fraction $1/\eta^*$ in the 3D WCA-HS system, with fitting $\xi=\xi_0[\eta^*/(\eta^*_0-\eta^*)]^\nu$ with $\eta^*_0=0.6033$ and a fitted $\nu=0.52$ for both $\xi_\Gamma$ and $\xi_4$.
     The vertical dashed lines indicate the mode-coupling temperature (or mode-coupling packing fraction).}
	\label{fig:correl} 
\end{figure}

We examined the spatial correlations of the slow particles and glass-like regions identified by the Grad-CAM field to characterize the dynamical and structural heterogeneity further. We first calculated the four-point radial correlation function of the slow particles, $g_4(r)$, as defined in equation~(\ref{eq:g4}) in Methods. As shown in Fig.~\ref{fig:g4}a, c, and e, the $g_4(r)$ curves for the three systems exhibited oscillatory peaks at characteristic interparticle distances, similar to those of the radial distribution function $g(r)$~\cite{Kob_1995KAM,Lacevic_2003chi4}. The peak envelope decayed toward zero with increasing distance $r$. The peak heights increased upon cooling or compression, indicating increasingly pronounced spatial correlations among the slow particles. Figure~\ref{fig:g4}b, d, and f show the corresponding radial correlation functions of the glass-like particles, $g_{\Gamma}(r)$ [equation~(\ref{eq:g_gamma}) in Methods]. These functions similarly exhibited oscillatory peaks that decayed with the distance. Their increasing peak heights upon cooling or compression indicate increasing spatial correlations in the Grad-CAM structural field.

We estimated the correlation lengths of the slow clusters, $\xi_4$, and glass-like structures, $\xi_\Gamma$, by fitting the envelopes of the decaying peaks of $g_4(r)$ and $g_{\Gamma}(r)$, respectively, to $f(r)=A\exp(-r/\xi)$~(where $A$ and $\xi$ are fitting parameters)~\cite{Lacevic_2002g4}. The resulting temperature or packing-fraction dependence of $\xi_4$ and $\xi_{\Gamma}$ is shown in Fig.~\ref{fig:correl}. 
The increase in $\xi_\Gamma$ was robust against the choice of the threshold that is used to define glass-like particles and whether $g_{\Gamma}(r)$ is scaled by $g(r)$ (Supplementary Figs.~9 and 10).
In the 3D KA model, $\xi_4$ increased upon cooling down to the mode-coupling temperature $T_{\mathrm{MCT}}$, and then saturated below $T_{\mathrm{MCT}}$ (blue symbols). This behavior is qualitatively consistent with that of the peak dynamic susceptibility $\chi^\mathrm{p}_4$~\cite{ortlieb_2023,Das_2022chi4,Liu_2024} (Supplementary Fig.~4b), reflecting the increased dynamical heterogeneity on approaching $T_{\mathrm{MCT}}$. 
In contrast, $\xi_\Gamma$ (purple symbols) continued to increase over the investigated temperature range, including below $T_{\mathrm{MCT}}$. The different low-temperature evolution of $\xi_4$ and $\xi_\Gamma$ may have contributed to the decrease in the structure--dynamics correlation coefficients shown in Fig.~\ref{fig:corref}a. 
The 3D WCA-HS system exhibited similar behavior.
In the 3D WAH mixture, both $\xi_4$ and $\xi_\Gamma$ increased monotonically with a decreasing temperature, including for $T<T_{\mathrm{MCT}}$. This is consistent with the enhanced structure--dynamics correlations shown in Fig.~\ref{fig:corref}b.

We fit $\xi_4$ and $\xi_\Gamma$ for $T>T_{\mathrm{MCT}}$ using
$\xi=\xi_0\{T_0/(T-T_0)\}^\nu$ with $\xi_0$ and $\nu$ as the fit parameters~(black dashed curves) to compare the temperature dependence of the two length scales. Here, $T_0$ is the extrapolation temperature obtained from Vogel--Fulcher--Tammann~(VFT) fits~\cite{Sengupta_2013chi4} to the structural relaxation time $\tau_\alpha$ (Supplementary Fig.~4). The fitted exponents $\nu$ for $\xi_4$ and $\xi_\Gamma$ were similar (see the caption of Fig.~\ref{fig:correl}), indicating that the two correlation lengths exhibited closely related temperature/fraction dependences over this range. These exponents are also comparable to previously reported values~\cite{Kawasaki_2007op2,Kawasaki_2011op2,Tanaka_2010op2,Tong_2018TT}.
We used this functional form only to compare the temperature or fraction dependences of the two lengths. Given the limited range of $\xi$ and the temperature, the present data could not establish whether either length diverged at $T_0$. Moreover, other functional forms may describe the data equally well.
Notably, in the 3D WCA-HS model, both $\xi_\Gamma$ and $\xi_4$ increased by less than a factor of two over the accessible range. Thus, the similar exponents indicate consistency, rather than a common origin. Whether $\xi_\Gamma$ continued to track $\xi_4$ deeper in the glassy regime remains unknown. Within these limits, the present results support an association between the increase in the structural correlations identified by Grad-CAM and the development of dynamical heterogeneity.

\begin{figure*} %
	\centering
	\includegraphics[width=0.9\textwidth]{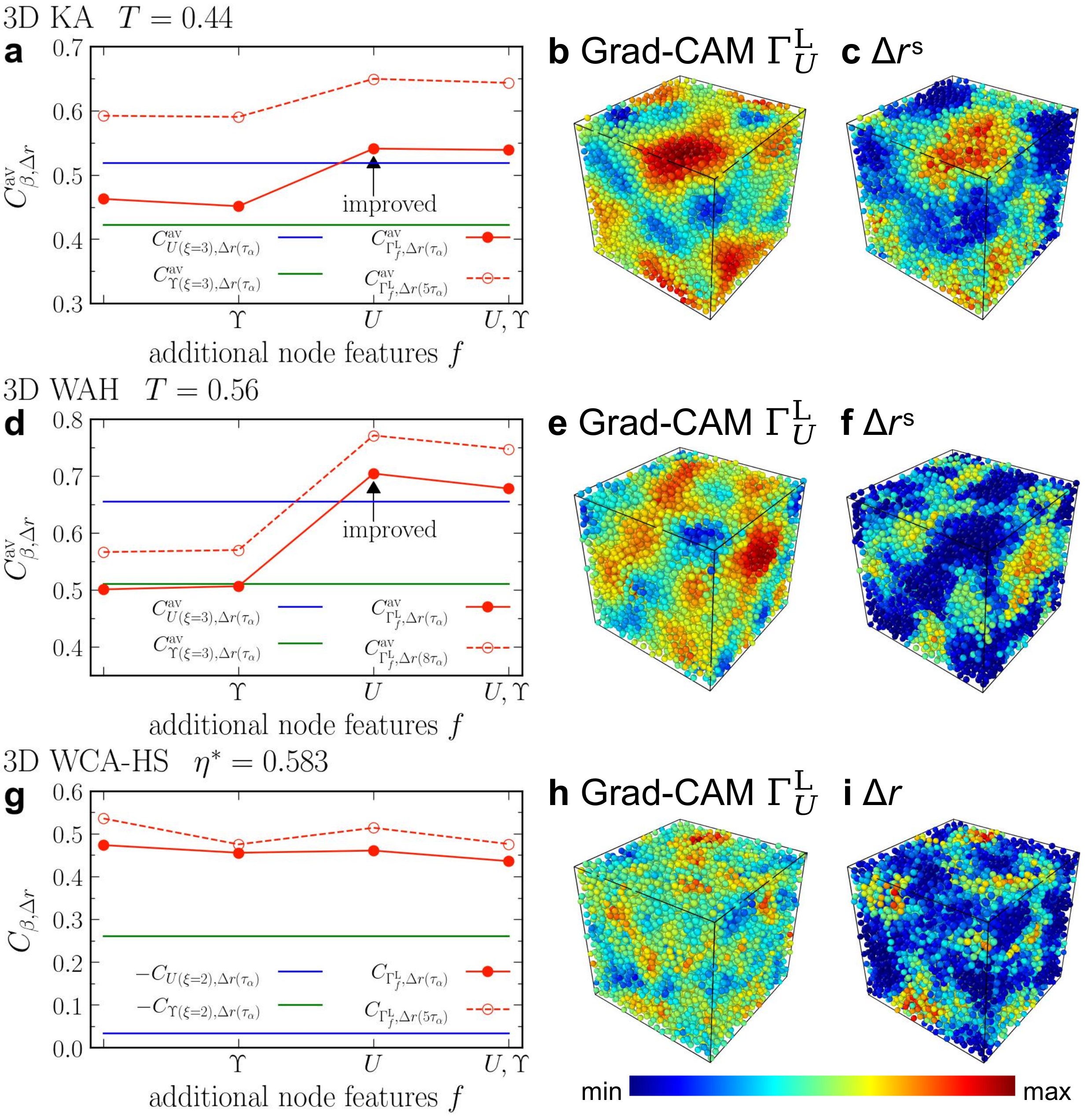}
	\caption{\textbf{Grad-CAM scores obtained from GNNs with different training ingredients.}
    \textbf{a} Pearson's correlation coefficients $C^{\mathrm{av}}_{\Gamma^{\mathrm{L}},\Delta r}$ between the dynamic propensity $\Delta r$ and Grad-CAM scores $\Gamma^{\mathrm{L}}_f$ extracted by the GNNs with different additional node features $f$ ($f\in\{\Upsilon, U\}$) at $T=0.44$ in the 3D KA model.
    The data on the far left indicate correlations with the original $\Gamma^{\mathrm{L}}$ without additional node features (trained only on the particle type).
    The solid and open circles denote correlations with $\Delta r$ over $\tau_\alpha$ and $5\tau_\alpha$, respectively, as in Fig.~\ref{fig:corref}.
    The correlation coefficients between dynamics and the coarse-grained potential $U$ (blue line) and between dynamics and the coarse-grained Voronoi volume $\Upsilon$ (green line) are shown as a reference. 
    Both $U$ and $\Upsilon$ were coarse-grained with a length of $\xi=3$, and their correlations with the dynamics showed almost the maximum values~(see Supplementary Fig.~11).
    \textbf{b} Snapshot of Grad-CAM scores $\Gamma_U^{\mathrm{L}}$ obtained from the GNN with training potential $U$, and (\textbf{C}) snapshot of the scaled dynamic propensity $\Delta r^\mathrm{s}$ over $\tau_\alpha$.
    \textbf{d}-\textbf{f} The same as in \textbf{a}-\textbf{c}, but the open circles indicate $\Delta r$ over $8\tau_\alpha$ for the 3D WAH mixture at $T=0.56$.
    \textbf{g}-\textbf{i} The same as in \textbf{a}-\textbf{c}, but shown as $-C_{U(\xi=2),\Delta r}$, $-C_{\gamma(\xi=2),\Delta r}$ and $C_{\Gamma^\mathrm L,\Delta r}$ in \textbf{g}, and the dynamic propensity $\Delta r$ in \textbf{i} for the 3D WCA-HS model at $\eta^*=0.583$.
    The values in the snapshots range from blue to red, as indicated by the color bar.
    The correlations between the Grad-CAM scores and dynamics for the aforementioned two systems improved when training $U$ [$C^\mathrm{av}_{\Gamma^{\mathrm{L}},\Delta r(\tau_\alpha)}=0.54$ and $C^\mathrm{av}_{\Gamma^{\mathrm{L}}_U,\Delta r(5\tau_\alpha)}=0.65$ for the 3D KA model, and $C^\mathrm{av}_{\Gamma^{\mathrm{L}},\Delta r(\tau_\alpha)}=0.70$ and $C^\mathrm{av}_{\Gamma^{\mathrm{L}},\Delta r(8\tau_\alpha)}=0.77$ for the 3D WAH model].
    However, the original Grad-CAM scores $\Gamma^{\mathrm{L}}$ showed the best performance for the 3D WCA-HS system.
    }
	\label{fig:gnn_improve} %
\end{figure*}

\subsection*{GNN with additional structural indicators.}
The GNN discussed above used the particle type as a node feature and particle positions to construct the graph representation.
The potential energy and Voronoi volume showed spatial correlations with the dynamics in the 3D KA and 3D WAH systems (Fig.~\ref{fig:snap3d}). Therefore, we examined whether including these quantities as additional node features improved the correlation between the Grad-CAM structural field and dynamics. Figure~\ref{fig:gnn_improve} shows the Pearson's correlation coefficients between the dynamic propensity $\Delta r$ and Grad-CAM scores $\Gamma^\mathrm{L}$ obtained from GNNs that were trained with different node features.

For the 3D KA system, the original GNN, which used only the particle type as the node feature, yielded Grad-CAM scores that correlated more strongly with the dynamics than the coarse-grained Voronoi volume $\Upsilon$~($C^\mathrm{av}_{\Gamma^{\mathrm{L}},\Delta r}>C^\mathrm{av}_{\gamma(\xi=3),\Delta r}$), but less strongly than the coarse-grained potential energy $U$~($C^\mathrm{av}_{\Gamma^{\mathrm{L}},\Delta r}<C^\mathrm{av}_{U(\xi=3),\Delta r}$). 
For the 3D WAH model, this correlation was comparable to that with the coarse-grained Voronoi volume $\Upsilon$ and lower than that with the coarse-grained potential energy $U$. 
Including $\Upsilon$ as an additional node feature resulted in minimal change in the correlations in either system. This result is consistent with $\Upsilon$ providing mainly local geometric information that is already accessible from the particle coordinates~\cite{Sharma_interpret2026}. 
In contrast, including $U$ substantially enhanced the correlation, by approximately $20\%$--$40\%$, and yielded correlations that were higher than $C^\mathrm{av}_{U~(\xi=3),\Delta r}$. 
The corresponding Grad-CAM snapshots based on the GNN with $U$ added for training are shown on the right of Fig.~\ref{fig:gnn_improve}.
Notably, in the 3D KA model at \(T=0.44\), the Grad-CAM scores obtained by training with the potential energy \(U\), denoted by \(\Gamma^\mathrm{L}_U\), exhibited a correlation coefficient of up to 0.65 with \(\Delta r(5\tau_\alpha)\). 
This value is comparable to those reported for some machine-learning models that were trained using dynamical information~\cite{Bapst_2020GNN,Boattini_2021ML}.
In the WAH system, \(C^\mathrm{av}_{\Gamma^\mathrm{L}_U,\Delta r(8\tau_\alpha)}\) reached almost 0.8 at \(T=0.56\), demonstrating a remarkably strong correlation and the powerful predictive capability of this GNN model.
However, the original Grad-CAM scores $\Gamma^\mathrm{L}$ for the 3D WCA-HS system, which were obtained without including either $U$ or $\Upsilon$ as node features, showed the strongest correlation with the dynamics. In this system, both conventional indicators, $U$ and $\Upsilon$, exhibited relatively weak correlations with the dynamic propensity.

These results indicate that the relative information provided by the geometric and energetic descriptors was system dependent. The particle coordinates alone yielded a structural field that was strongly associated with dynamical heterogeneity across the three models. In the attractive 3D KA and 3D WAH systems, the local energetic information also contributed to this association, whereas in the purely repulsive 3D WCA-HS system, the coordinate-based GNN provided the strongest structure--dynamics correlation.

\section*{Discussion}\label{sec:discu}

In this study, we have shown that an increasing structural length scale that is associated with dynamical heterogeneity in supercooled liquids can be extracted from static configurations alone, without using dynamical information for training or prescribing a coarse-graining length. A GNN that was trained only to distinguish liquid and glass configurations, combined with Grad-CAM, yielded a particle-resolved structural field (Fig.~\ref{fig:gnn}) with liquid-like and glass-like regions that corresponded to mobile and immobile regions, respectively, particularly at low temperatures or high packing fractions (Figs.~\ref{fig:snap3d} and \ref{fig:snap3dhs}). As the network did not receive dynamical observables, this correspondence was not imposed by supervision. The structure--dynamics correlation strengthened upon cooling or compression over the investigated range (Fig.~\ref{fig:corref}). Moreover, a network that was trained on systems of $N=1000$ particles could be applied directly, without retraining, to systems of up to $N=40000$ particles (Fig.~\ref{fig:extend}). This is consistent with the structural field being dominated by local and mesoscopic correlations, as opposed to system-size-specific features. This transferability enables analyses of systems that are sufficiently large to resolve increasing mesoscopic correlations at a modest training cost.

The central result is that the size of the glass-like regions identified by the Grad-CAM field, $\xi_\Gamma$, increased upon cooling or compression and followed the dynamical correlation length $\xi_4$ above the mode-coupling crossover in all three models, with closely similar effective exponents (Fig.~\ref{fig:correl}). In this case, $\xi_\Gamma$ was defined by a single, state-independent criterion, directly analogous to the fixed mobility threshold that defines $\xi_4$. Therefore, it measured the spatial extent of regions that shared the same structural characteristics at all state points. Its temperature dependence was robust against the choice of threshold as well as against the normalization of $g_\Gamma(r)$ by $g(r)$ (Supplementary Figs.~9 and 10).
To the best of our knowledge, such an increasing static length that tracks $\xi_4$ has not previously been obtained without any dynamical input. In a previous unsupervised approach, the correlation length of the learned order parameter remained essentially constant upon supercooling, and an increasing length appeared only following local averaging optimized against the dynamic propensity~\cite{Boattini_2020ML}. This difference also reflects the distinct quantities measured.

The comparison of the GNNs that were trained with different node features showed that the local information most relevant to the dynamics was dependent on the interaction (Fig.~\ref{fig:gnn_improve}). In the attractive 3D KA and 3D WAH systems, adding the local potential energy $U$ increased the correlation around $\tau_\alpha$ to $0.54$ and $0.70$, respectively, whereas adding the Voronoi volume $\Upsilon$ produced minimal change. This is consistent with the geometric information already accessible from the particle coordinates. In the purely repulsive 3D WCA-HS system, the coordinate-based GNN yielded the strongest correlation, whereas potential energy $U$ and the Voronoi volume $\Upsilon$ correlated only weakly with the dynamics; the sign of the relation between the Voronoi volume and mobility even differed between the 3D KA and 3D WCA-HS systems (Figs.~\ref{fig:snap3d} and \ref{fig:snap3dhs}). Despite these differences in the relevant local information, the same architecture and training protocol yielded an increasing structural length that accompanied dynamical heterogeneity in all three systems. Therefore, as opposed to a specific local motif, the common factor across these glass formers is the spatial organization of structurally distinct regions, the extent of which increases together with that of dynamical heterogeneity. The framework identifies this organization without presupposing the importance of the local features. Thus, it can detect this growth even when no single conventional descriptor performs consistently across systems.

The present framework opens several directions. First, a systematic comparison with other machine-learning approaches would clarify the dependence of the extracted length on the input information, network architecture, and definition of the correlation length. Second, a comparison with the point-to-set length and with known structural orders in the same systems would clarify the physical meaning of $\xi_\Gamma$ and whether the network captures the same type of structural information across systems. Such static lengths may behave differently from one another and from dynamical lengths~\cite{Hocky_2012,Charbonneau_2013,Hocky_2014,Russo_2015}. In particular, static lengths have been reported to increase weakly in HS glass formers~\cite{Charbonneau_2013}, which motivates a direct comparison in the 3D WCA-HS system. Third, whether the increase in dynamical correlations is driven by structure or by kinetic constraints is not decided by the present correlations alone. The present field can be used to determine the relation to dynamical facilitation, in which mobility spreads through kinetic constraints~\cite{Garrahan_2002,Chandler_2010g,Keys_2011}, and which plays a central role in deeply supercooled liquids~\cite{Scalliet_2022,Guiselin_2022}. Our results are compatible with a combined picture in which the local structure biases the location of relaxation initiation while facilitation governs the spread of mobility~\cite{ortlieb_2023}. A time-resolved analysis of whether relaxation initiates in and propagates from liquid-like regions would clarify whether the departure of $\xi_\Gamma$ from $\xi_4$ below the mode-coupling crossover reflects a change in the dominant relaxation mechanism.

In summary, we established a general scheme for identifying increasing static structural signatures that are associated with dynamical heterogeneity in the super-Arrhenius regime, even when the relevant local structures differ across glass formers. A GNN that is trained only to distinguish liquid and glass configurations, without dynamical information or a prescribed coarse-graining length, yields a particle-resolved structural field in which the glass-like regions increase upon cooling or compression in parallel with the dynamical correlation length. The trained network is directly applicable to substantially larger systems than those used for training and is robust across the training conditions examined. Therefore, the framework provides a scalable route to characterizing structural correlations and testing their relation to the dynamics. This includes glass-forming systems for which no effective structural descriptor has yet been identified~\cite{Berthier_2009hm, Ikeda_2011gcm} and more complex disordered systems such as dense cell collectives.

\section*{Methods}~\label{sec:method}
\vspace{-8ex}
\subsection*{Simulations of the KA model.}~\label{sec:md}
We carried out molecular dynamics simulations of the KA mixture in both $2$D and $3$D spaces~\cite{Kob_1995KAM}.
The particle interactions are described by the following Lennard--Jones potential:
\begin{equation}\label{eq:energy}
\begin{aligned}
&\Phi(r) = \Phi^*(r)-\Phi^*(r_{ij}^c)-(r-r_{ij}^c) \left. \frac{\mathrm{d}\Phi^*}{\mathrm{d}r}\right|_{r = r_{ij}^c},\\
&\Phi^*(r) = 4\epsilon_{ij} \left[ \left( \frac{\sigma_{ij}}{r} \right)^{12} - \left( \frac{\sigma_{ij}}{r} \right)^{6} \right],
\end{aligned}
\end{equation}
where $r$ is the distance between particles $i$ and $j$.
The potential is smoothed at the cutoff distance $r_{ij}^c = 2.5\sigma_{ij}$ up to the first derivative of the potential~\cite{Mizuno_2016}.
The composition of species A and B is $N_\mathrm{A}:N_\mathrm{B}=65:35$ for the 2D system and $80:20$ for the 3D system.
The parameters $\sigma_{ij}$ and $\epsilon_{ij}$ were defined as $\sigma_\mathrm{AA} = \sigma$, $\sigma_\mathrm{BB} = 0.88\sigma$, $\sigma_\mathrm{AB} = 0.8\sigma$, and $\epsilon_\mathrm{AA} = \epsilon$, $\epsilon_\mathrm{BB} = 0.5\epsilon$, $\epsilon_\mathrm{AB} = 1.5\epsilon$, respectively.
The mass of both species $\mathrm{A}$ and $\mathrm{B}$ was set to $m_\mathrm{A}=m_\mathrm{B}=m$.
$\sigma$, $\epsilon$, and $\sqrt{m \sigma^2/\epsilon}$ are the units of length, energy, and time, respectively~(which are applicable to other systems).
The number density $\rho$($\rho=N/L^2$ for 2D and $N/L^3$ for 3D) was set to $1.2$, where $N$ is the particle number and $L$ is the box length.
We considered periodic boundary conditions in all directions based on the open-source software LAMMPS~\cite{Plimpton_1995,lammps} for all the systems.

\subsection*{Simulations of the 3D WAH model.}~\label{sec:md-wah}
We also performed molecular dynamics simulations of the 3D WAH model~\cite{wah_1991,Lacevic_2003chi4}, in which particles interacted through the Lennard--Jones potential defined in equation~(\ref{eq:energy}).
The main parameters were defined as follows: $\rho=1.296$, $N_\mathrm{A}:N_\mathrm{B}=50:50$, $r_{ij}^c = 2.5\sigma_{ij}$, $\sigma_\mathrm{AA} = \sigma$, $\sigma_\mathrm{BB} = 5/6\,\sigma$, $\sigma_\mathrm{AB} = (\sigma_\mathrm{AA} +\sigma_\mathrm{BB})/2$, $\epsilon_\mathrm{AA} =\epsilon_\mathrm{BB}=\epsilon_\mathrm{AB} = \epsilon$, $m_\mathrm{A}=2m$, and $m_\mathrm{B}=m$.

\subsection*{Simulations of the 3D WCA-HS model.}~\label{sec:md-hs}
We employed the WCA repulsive potential for the HS system~\cite{wca_1971}:
\begin{equation}\label{eq:energy_wca}
\begin{aligned}
&\Phi(r) = \begin{cases} 
4\epsilon_{ij} \left[ \left( \frac{\sigma_{ij}}{r} \right)^{12} - \left( \frac{\sigma_{ij}}{r} \right)^{6} + \frac{1}{4} \right] \hspace{0.1cm} r<2^{1/6}\sigma_{ij}\\
0 \hspace{4.1cm} r\geq 2^{1/6}\sigma_{ij}.
\end{cases}
\end{aligned}
\end{equation}
We fixed the temperature to $T=0.1$ and controlled the packing fraction of this system.
The effective packing fraction is expressed as $\eta^*=\pi\sum_{j=1}^N(\sigma^*_j)^3/(6L^3)$, where $\sigma^*_j$ indicates the effective diameter of particle $j$.
$\sigma^*$ is determined with $\Phi(\sigma^*)=k_BT$, which is known as the Boltzmann criterion~\cite{eff-d_2022}.
The main parameters followed those of a previous paper~\cite{Boattini_2020ML}: $N_\mathrm{A}:N_\mathrm{B}=30:70$, $\sigma_\mathrm{AA} = \sigma$, $\sigma_\mathrm{BB} = 0.85\,\sigma$, $\sigma_\mathrm{AB} = (\sigma_\mathrm{AA} +\sigma_\mathrm{BB})/2$, $\epsilon_\mathrm{AA} =\epsilon_\mathrm{BB}=\epsilon_\mathrm{AB} = \epsilon$, and $m_\mathrm{A}=m_\mathrm{B}=m$.

\subsection*{Sample preparation.}~\label{sec:sample}
We explain the sample preparation for all the systems.
We prepared $N_\mathrm{s}=1000$ liquid samples and $N_\mathrm{s}=1000$ glass samples, each containing $N=1000$ particles, for the machine-learning training input.

Initially, random configurations in the liquid state were generated and equilibrated at a high temperature of $T = 2.0$ for the 3D KA and 3D WAH systems or at a low packing fraction of $\eta^*=0.370$ for the 3D WCA-HS model.
Thereafter, the samples were cooled to $T_\mathrm{L}$~($T_\mathrm{L}=1.0$ for the 3D KA and $T_\mathrm{L}=1.8$ for the 3D WAH models) or compressed to $\eta^*_\mathrm{L}$ ($\eta^*_\mathrm{L}=0.493$ for the WCA-HS model), and the liquid state was equilibrated.
$T_\mathrm{L}$ is near or above the onset temperature, and $\eta^*_\mathrm{L}$ is below the onset packing fraction, where almost no plateau is observed in the self-intermediate scattering function $F_\mathrm{s}(q,t)$.
To obtain the glass configurations, we continued cooling toward $T_\mathrm{G}=0.05$ with a constant cooling rate of $v_\mathrm{c}=\mathrm{d}T/\mathrm{d}t = 10^{-5}$ for the 3D KA and 3D WAH models, and compressing toward $\eta^*_\mathrm{G}=0.70$ with a rate of $v_\mathrm{c}=\mathrm{d}\eta^*/\mathrm{d}t = 10^{-5}$ for the 3D WCA-HS system.
These parameters, including $N_\mathrm{s}$, $T_{\rm G}$ ($\eta_{\rm G}^*$), $T_{\rm L}$ ($\eta_{\rm L}^*$), $v_{\rm c}$, and $N$, were varied in additional training runs to confirm the robustness of our results~(Supplementary Figs.~7 and~8).

We focused on a wide range of temperatures~(or packing fractions) to investigate the supercooled-liquid states: $0.42 \le T \le 1.0 $ for the 3D KA system, $0.56 \le T \le 1.0 $ for the 3D WAH system, and $0.530 \le \eta^*\le 0.586 $ for the 3D WCA-HS system.
A relatively larger system with particle number $N=10000$ was employed for the main analysis~(see Fig.~\ref{fig:extend}).
A total of $50$ independent samples were prepared to calculate the ensemble average of the physical quantities.
Starting from the initial equilibrium configurations at $T_\mathrm{L}$~($\eta^*_\mathrm{L}$), we cooled~(compressed) the systems to each target temperature~(packing fraction) and subsequently fully equilibrated them through NVT simulations with the Nos$\acute{\text{e}}$--Hoover thermostat~\cite{Evans_2008}.
We utilized equilibrium configurations in this study.
The inherent structures of these configurations have been confirmed to show qualitatively similar behavior~(see Supplementary Fig.~6 and Supplementary Table~2).

\subsection*{Classification of glass and liquid using a GNN.}~\label{sec:gnn}
The classification based only on particle configurations using the GNN is described in this section.
As mentioned previously, we mainly generated two datasets of liquid configurations at $T_\mathrm{L}$~($\eta^*_\mathrm{L}$) and glassy solid configurations at $T_\mathrm{G}$~($\eta^*_\mathrm{G}$) with a size of $N=1000$ for the 3D systems.
We utilized $N_\mathrm{s}=1000$ samples for each state, with $800$ samples for training and $200$ samples for validation~(2000 samples in total).

As shown in Fig.~\ref{fig:gnn}, the particle configurations are initially converted into graphs as the input to the network.
In these graphs, each particle is considered as one node, with the particle type as the node feature. 
Other information, such as the potential energy, is also included in the node features for comparison~(see Fig.~\ref{fig:gnn_improve}).
The number of features $n$ is the dimension of the node features.
All results presented are based on the GNN using only the particle type as the node feature ($n=1$), unless otherwise stated.
For each particle pair $(i, j)$ within a distance of $2$, two edges are created in opposite directions: one from $i$ to $j$ and the other from $j$ to $i$. 
The relative distance vector between connected particles is used as the feature of the corresponding edge, with the same dimension as the dimension $d$ of the system. 
For example, in 3D, the edge feature from particle $i$ to particle $j$ is expressed as ($r_{ij}^x$, $r_{ij}^y$, $r_{ij}^z$), where 
$r_{ij}^{\mu}=r_j^{\mu}-r_i^{\mu}$~($\mu=x,y,z$) is calculated under periodic boundary conditions and $r_i^{\mu}$ denotes the coordinate of particle $i$ along the $\mu$ direction.
Our previously proposed CNN approach~\cite{Oyama_2023, Liu_2024, Liu_2025} using grid data based on the particle density could not capture information regarding the particle type, whereas the current GNN method can incorporate such messages.

The GNN architecture is similar to those in previous works~\cite{Bapst_2020GNN, Shiba_2023GNN}.
First, a two-layer multi-layer perceptron (MLP) is used as an encoder to assign 64-dimensional~($64$D) features for each node and edge in the input graph.
Subsequently, an additional two-layer MLP is employed to update these features iteratively. 
The edge features (see the red edges as an example in the ``Edge update" frame) are updated at each iteration based on their features, together with those of the two connected nodes (highlighted in yellow), following which the node features (refer to the red node in the ``Node update" frame) are updated using the features of the given node and the edges toward them (highlighted in yellow).
The encoded features of nodes and edges are also input for their updates to provide stable learning.
Up to this point, the learning architecture is identical to that used in Ref.~\cite{Shiba_2023GNN}.
In our framework, the same message-passing rule is repeated $N_{\mathrm{ite}}=5$ times.
Accurate classification can be achieved when \(N_{\mathrm{ite}} \geq 3\).
We verified that the results are optimal with $N_{\mathrm{ite}}=5$ and do not change qualitatively under small variations in $N_{\mathrm{ite}}$~\cite{Shiba_2023GNN}~(see Supplementary Fig.~5).

We add a graph convolution operation to the final node embedding following message passing, which generates $N_\mathrm{f}=32$ feature maps from the node representations, to apply Grad-CAM to the GNN.
Subsequently, global average pooling is performed on each feature map and $N_\mathrm{f}=32$ values are obtained.
We confirmed that the results with pooling are highly similar to those without pooling.
Owing to this procedure~(highlighted by the purple shaded region in Fig.~\ref{fig:gnn}), the network can be applied to systems of any size after being trained on the dataset with a single size.
In this case, the GNN is trained using configurations with $N=1000$, but it can be applied to systems with different sizes of $N=1000$, $N=4000$, $10000$, and $40000$~(Fig.~\ref{fig:extend}).
Further results are provided in Supplementary Figs.~7 and 8.
This process significantly reduces the training cost, which is one of the main advantages and highlights of our protocol.

Subsequently, the $N_\mathrm{f}=32$ values are fed into a three-layer MLP, with the number of output values for each layer shown below.
The graph convolution and all layers in the MLP are followed by rectified linear unit (ReLU) activations, except for the final layer of the network.
The target output of the network is the two classification scores, $S_C$, where $ C$ denotes liquid ($C=\mathrm L$) or glass ($ C=\mathrm G$).
Subsequently, the softmax function is applied to these scores ($0\leq P_C \leq 1$; $P_\mathrm {G}+P_\mathrm{L}=1$) to obtain the probability $P_C$ of the configuration for each class $C$.
Refer to the supplemental material of Ref.~\cite{Oyama_2023} for further details on the loss function and optimization.
 
The earliest learning epoch at which the classification accuracy exceeded 99\% (i.e., almost 100\%) and the loss fell below 10\% on both the training and validation sets was selected for analysis.
Small changes in the number of epochs did not qualitatively affect the results. However, excessively large epochs should be avoided to prevent overfitting.
The optimal epochs varied across systems and training datasets. 
For example, the epoch numbers were fixed at 100 for the 3D KA mixture, 216 for the 3D WAH model, and 24 for the 3D WCA-HS system.
We confirmed that the GNN results with varying training data conditions, such as the number of training samples $N_\mathrm{s}$, training points $T_\mathrm G$ and $T_\mathrm L$, cooling or compressing rate $v_\mathrm{c}$, and training system size $N$, were qualitatively the same as those based on the initial training set (see Supplementary Figs.~7 and 8).
Notably, the GNN was trained on bare static data, without any dynamic information, but it can be applied to predict dynamical heterogeneity in supercooled liquids.

\subsection*{Extraction of structural indicators using Grad-CAM.}~\label{sec:gradcam}
We explain the extraction of the local structural features using the Grad-CAM method~\cite{Selvaraju_2020gradcam, Font-Clos_2022}.
Grad-CAM quantifies the influence of the feature maps $\boldsymbol{A}=\{A_i^k\}_{i=1,2,..., N}^{k=1,2,..., N_\mathrm{f}}$ on the resulting classification score $S_C$ of the GNN. Subsequently, it extracts the significant structural features as Grad-CAM scores, as follows:

\begin{equation}\label{eq:gradcam}
\begin{aligned}
    \alpha_{k}^C  & = \frac{1}{N}\sum_{i}^N \frac{\partial S_C}{\partial A_i^k}, \\
    \Gamma_i^{C} & = \sum_{k}^{N_\mathrm{f}} \alpha_{k}^C A_i^k, 
\end{aligned}
\end{equation}
where $A_i^k$ is the $i$th node in the $k$th feature map $\boldsymbol{A}^k=\{A_i^k\}_{i=1,2,...,N}$, $\alpha_{k}^C$ evaluates the contribution of the $k$th feature map $\boldsymbol{A}^k$ to the classification score $S_C$ by differentiating $S_C$ with respect to the feature map $A_i^k$, and $\Gamma_i^{C}$ indicates the Grad-CAM score of particle $i$ (the $i$th node) for class $C$, which is the weighted summation of the feature maps. 

The clear interpretability of the Grad-CAM scores is a strong advantage of this method.
Specifically, larger values of $\Gamma^\mathrm L$ indicate more liquid-like structures, whereas smaller values indicate more glass-like (less liquid-like) structures.
As the Grad-CAM scores $\Gamma^\mathrm {G}$ for glass contained largely complementary information to $\Gamma^\mathrm{L}$, the discussion proceeds primarily using one score, namely the liquid score $\Gamma^\mathrm {L}$.
\subsection*{Dynamic propensity.}~\label{sec:dp}
In this study, we calculated the dynamic propensity, which is widely used to investigate dynamical heterogeneities that are purely induced by structures~\cite{Widmer-Cooper_2004iso,Widmer-Cooper_2006iso,Widmer-Cooper_2008iso}.
The dynamic propensity is the average of the particle displacements based on an isoconfigurational ensemble: multiple trajectories from the same equilibrium configuration are computed with different initial velocities that follow a Maxwell--Boltzmann distribution.
The dynamic propensity $\Delta r$ over the timescale $t$ is defined as
\begin{equation} \label{eq:propensity}
 \Delta r_i(t)=\left< \left| \boldsymbol{r}_i(t)-\boldsymbol{r}_i(0) \right| \right>_\text{iso},
\end{equation}
where $\boldsymbol{r}_i(t)$ indicates the position of particle $i$ at time $t$, and $\left< \right>_\text{iso}$ denotes the average over the isoconfigurational ensemble.
A higher value of $\Delta r_i$ means that particle $i$ is mobile and a lower value means that it is immobile.
Previous studies have reported that structures show clear correlations with dynamics around the relaxation time $\tau_{\alpha}$~\cite{Boattini_2020ML,Susana_2020dp, Pezzicoli_2024GNN}. Thus, we mainly studied the dynamic propensity $\Delta r(\tau_\alpha)$ over the $\alpha$ relaxation time $\tau_\alpha$~(unless otherwise specified, $\Delta r$ denotes $\Delta r(\tau_{\alpha})$).
$\tau_\alpha$ was examined from the self-intermediate scattering function $F_\mathrm{s}(q,t)$ (see Supplementary Figs.~3 and 4 for further details).
We used 50 trajectories for the isoconfigurational average, which was sufficient to obtain accurate ensemble averages~\cite{Liu_2024}.

\subsection*{Radial correlation functions.}~\label{sec:sptial_c}
We studied the four-point functions $g_4(r)$ that involve only slow particles~\cite{Lacevic_2002g4,Lacevic_2003chi4} to measure the radial correlations for clusters of slow particles quantitatively.
First, we studied the dynamics with an overlap function:
\begin{equation}\label{eq:overlap}
 Q_4(t)=\sum_{i}^N \omega^4_i(t),
\end{equation}
with 
\begin{equation} \label{eq:w_dy}
	\omega_i^4(t)=\begin{cases}
        0~~|\mathbf{r}_i(t)-\mathbf{r}_i(0)|> \Delta r_\mathrm{c}\\
        1~~|\mathbf{r}_i(t)-\mathbf{r}_i(0)| \leq \Delta r_\mathrm{c}
        \end{cases},
\end{equation}
where $\Delta r_\mathrm{c}=0.3$ is the mobility threshold: $\omega_i^4=0$ indicates fast dynamics and $\omega_i^4=1$ indicates slow dynamics.

Then, $g_4(r)$ over a timescale of $\tau_\alpha$ is defined via the following function:
\begin{equation}\label{eq:g4}
\begin{split}
	g_4(r)={}&\frac{1}{N\rho}\sum_{l \neq k}^{N}
	\left\langle \omega_l^4(\tau_\alpha)\omega_k^4(\tau_\alpha)
	\delta\!\left(\mathbf{r}_l(0)-\mathbf{r}_k(0)-\mathbf{r}\right)\right\rangle \\
	&-\left\langle \frac{1}{N}Q_4(\tau_\alpha) \right\rangle^2,
\end{split}
\end{equation}
where $\left< \right>$ denotes the ensemble average.

We generalized the radial correlation functions for static structures and defined $g_{\Gamma}(r)$ for glass-like particles characterized by the Grad-CAM scores $\Gamma^\mathrm L$, as follows:

\begin{equation}\label{eq:g_gamma}
\begin{split}
	g_{\Gamma}(r)={}&\frac{1}{N\rho}\sum_{l \neq k}^{N}
	\left\langle \omega_l^\Gamma \omega_k^\Gamma
	\delta\!\left(\mathbf{r}_l-\mathbf{r}_k-\mathbf{r}\right)\right\rangle 
	-\left\langle \frac{1}{N}Q_\Gamma \right\rangle^2,
\end{split}
\end{equation}
with
\begin{equation}\label{eq:w_gamma}
	\omega_i^\Gamma=\begin{cases}
        0~~\Gamma^\mathrm L_i>\Gamma^\mathrm L_\mathrm{c}\\
        1~~\Gamma^\mathrm L_i \leq \Gamma^\mathrm L_\mathrm{c}
        \end{cases}, 
\end{equation}
where $\mathbf r_i$ is the position of particle $i$ and $Q_\Gamma = \sum_i^N \omega_i^\Gamma$ is the summation of the indicator $\omega^\Gamma$. $\omega_i^\Gamma = 0$ indicates the liquid-like state and $\omega_i^\Gamma = 1$ indicates the glass-like state. The threshold is defined as $\Gamma^\mathrm{L}_c = \Gamma^\mathrm{L}_\mathrm{min} + \lambda\,(\Gamma^\mathrm{L}_\mathrm{max} - \Gamma^\mathrm{L}_\mathrm{min})$ with $\lambda = 0.5$, where $\Gamma^\mathrm{L}_\mathrm{min}$ and $\Gamma^\mathrm{L}_\mathrm{max}$ are the minimum and maximum values over all temperatures $T_\mathrm{G} \le T \le T_\mathrm{L}$~(or all fractions $\eta^*_\mathrm{L} \le \eta^* \le \eta^*_\mathrm G$), respectively. We used a single temperature-independent threshold so that glass-like regions were defined by the same absolute criterion for all state points, analogous to the fixed mobility threshold $\Delta r_c$ used for $g_4(r)$. 
Therefore, the resulting length $\xi_\Gamma$~(see Figs.~\ref{fig:g4} and ~\ref{fig:correl}) characterized the spatial extent of the glass-like regions, rather than the correlation of fluctuations about the mean at each state point. We confirmed that the temperature dependence of $\xi_\Gamma$ remained essentially unchanged for $\lambda = 0.4$ and $0.6$ (Supplementary Fig.~9). 

\subsection*{Correlation coefficients.}~\label{sec:cor}
We studied Pearson's correlation coefficient to measure the spatial correlation between the structures and dynamic propensity $\Delta r$ within a configuration. Pearson’s correlation coefficient is defined as
\begin{equation}\label{eq:cor} 
C_{S,~\Delta r}=\frac{\sum_i^N (S_i- \langle S \rangle_\mathrm {p})(\Delta r_i- \langle \Delta r\rangle_\mathrm {p})}{\sqrt{\sum_i^N  (S_i- \langle S \rangle_\mathrm {p})^2}\sqrt{\sum_i^N  (\Delta r_i-\langle \Delta r\rangle_\mathrm {p})^2}}, 
\end{equation}
where $S$ indicates the structural indicators, such as the Grad-CAM scores $\Gamma^\mathrm L$, potential energy $U$, and Voronoi volume $\Upsilon$, whereas $\langle X \rangle_\mathrm {p}=\frac{1}{N}\sum_i^N X_i$ is the average value of $X$ ($X=S,\,\Delta r$) over the particles in this configuration.
A larger value of $C_{S,~\Delta r}$ suggests stronger correlations between the structures and dynamics in the configuration.

The correlations calculated for all particles in the 3D KA and 3D WAH models were lower than those obtained for individual particle species, as shown in Supplementary Table~1 and Supplementary Fig.~2. 
Thus, we separately calculated $C^\mathrm{A}_{S,~\Delta r}$ for A-particles and $C^\mathrm{B}_{S,~\Delta r}$ for B-particles~\cite{Boattini_2020ML,Paret_2020ML}, and then obtained the species-weighted average values (shown in Figs.~\ref{fig:corref}a, b, and~\ref{fig:gnn_improve}a, d): 
\begin{equation}\label{eq:corav}
C^{\mathrm{av}}_{S,~\Delta r} = (N_\mathrm{A} C^\mathrm{A}_{S,~\Delta r}+N_\mathrm{B} C^\mathrm{B}_{S,~\Delta r})/N.
\end{equation}

For the snapshots in Fig.~\ref{fig:snap3d}a, h and Fig.~\ref{fig:gnn_improve}c, f, we used the scaled dynamic propensity $\Delta r^{\mathrm{s}}_i=\Delta r_i/ \langle \Delta r_{\alpha(i)} \rangle$, where $\langle \Delta r_{\alpha(i)} \rangle$ is the average over species $\alpha$ to which particle $i$ belongs.
The value of $C_{\Gamma^\mathrm L,~\Delta r^\mathrm{s}}$ has been confirmed to be comparable to that of $C^{\mathrm{av}}_{\Gamma^\mathrm L,~\Delta r}$~(see the data in Supplementary Table~1). 
We directly used $C_{S,~\Delta r}$ for the correlation coefficients in the 3D WCA-HS in Fig.~\ref{fig:corref}c and Fig.~\ref{fig:gnn_improve}g and $\Delta r$ for the snapshots in Fig.~\ref{fig:snap3dhs}d, h and Fig.~\ref{fig:gnn_improve}i.
In addition, we confirmed that $C_{\Gamma^\mathrm L,~\Delta r}$ was close to $C^{\mathrm{av}}_{\Gamma^\mathrm L,~\Delta r}$ in the 3D WCA-HS model~(Supplementary Table~1).

\subsection*{Voronoi volume.}~\label{sec:vv}
In addition to the Grad-CAM scores, we analyzed the volume of the Voronoi cell associated with each particle.
The Voronoi cell ${\cal V}_i$ for particle $i$ is defined by the following equation:
\begin{align}\label{eq:voronoi}
{\cal V}_i = \lbrace n|\mathrm{L}(\mathbf n,\mathbf r_i)\le \mathrm{L}(\mathbf n,\mathbf r_j), j\ne i\rbrace,
\end{align}
where $\mathrm{L}(\mathbf p,\mathbf q)$ is a function for calculating the Euclidean distance between two points $\mathbf p$ and $\mathbf q$.
Point $\mathbf n$ is arbitrary and independent of the particle positions.
The volume $\Upsilon_i$ of the Voronoi cell ${\cal V}_i$ can characterize the degree of the local free volume, with a higher Voronoi volume suggesting a larger free volume.

\subsection*{Coarse graining of order parameters.}~\label{sec:cg}
As described previously, we studied the Grad-CAM scores from CNN $\Gamma_\mathrm{CNN}^\mathrm{L}$, potential energy $U$, and Voronoi volume $\Upsilon$ upon coarse-graining.
The coarse-graining procedure for these structural order parameters is performed as follows~\cite{Berthier_2007iso}:
\begin{equation} \label{eq:CG}
S_i(\xi)=\frac{\sum_{j}S^\mathrm{b}_j\mathrm{P}(|\boldsymbol{r}_j-\boldsymbol{r}_i|)}{\sum_{j}\mathrm{P}(|\boldsymbol{r}_j-\boldsymbol{r}_i|)},
\end{equation}
where $S$ represents $\Gamma_\mathrm{CNN}^\mathrm{L}$, $U$, and $\Upsilon$, whereas $S^\mathrm{b}_i$ and $S_i$ denote the particle-based bare value and coarse-grained value for particle $i$, respectively.
In addition, $\xi$ is the coarse-graining length scale. 
We employed the exponential function $\mathrm{P}(r)=\exp(-r/\xi)$ as our coarse-graining kernel~\cite{Tong_2018TT, Tong_2019TT}.
The summation $\sum_j$ considers all particles within a distance of $\xi$ from reference particle $i$, including $i$ itself.

\section*{Data availability}
The data that support the results in this paper are available at [https://doi.org/10.5281/zenodo.24002235].

\section*{Code availability}
The codes associated with this research are available from the corresponding author upon reasonable request.


\bibliographystyle{naturemag}
\bibliography{main} 


\section*{Acknowledgments}
We thank Dr. Lukas Fischer for his critical reading of the manuscript.
This work was supported by the JST FOREST Program (Grant No. JPMJFR212T), 
the AMED Moonshot Program (Grant No. JP22zf0127009), JSPS KAKENHI 
(Grant No. JP24H02203), and the Takeda Science Foundation.

\section*{Author contributions}
N.O. and T.K. designed and supervised the research.
All authors contributed to the Methodology.
M.L. and D.N. developed the software.
M.L. performed the simulations and data analysis. 
M.L. and T.K. wrote the original draft.
All authors reviewed and edited the paper.

\section*{Competing interests}
The authors declare no competing interests.

\clearpage

\begin{center}
{\bfseries Supplementary Information for\\
Revealing the growth of structural length in glass-forming liquids without using dynamical information}\\[0.7em]

Min~Liu,
Daisuke~Nishigaito,
Yusuke~Hara,
Norihiro~Oyama,
Takeshi~Kawasaki\\[2em]
\end{center}

\makeatletter
\renewcommand{\thefigure}{\arabic{figure}}
\renewcommand{\fnum@figure}{\textbf{Supplementary Figure~\thefigure:}}
\makeatother

\makeatletter
\renewcommand{\thetable}{\arabic{table}}
\renewcommand{\fnum@table}{\textbf{Supplementary Table~\thetable:}}
\makeatother

\titleformat{\subsection}[block]
  {\normalfont\normalsize\bfseries\centering}
  {\thesubsection}
  {1em}{}

\titlespacing*{\subsection}
  {0pt}
  {4ex}
  {2ex}
  
\renewcommand{\thepage}{S\arabic{page}}

\setcounter{subsection}{0} 
\setcounter{section}{0} 
\setcounter{figure}{0}
\setcounter{table}{0}
\setcounter{equation}{0}
\setcounter{page}{1} 

\subsection*{Snapshots of 2D KA model}
In the 2D KA mixture, pronounced structural heterogeneity was observed at a low temperature of $T = 0.38$, with liquid-like regions coexisting with glass-like regions, as shown in Supplementary Figure~\ref{fig:sup-snap2d}a.
When comparing with the snapshot of the dynamic propensity $\Delta r$ in Supplementary Figure~\ref{fig:sup-snap2d}b, we observed strong spatial correlations between them. 
More liquid-like structures tended to be associated with more mobile particles, whereas more glass-like structures were associated with more immobile particles.
These results demonstrate that the static structural features extracted by the GNN play a significant role in governing dynamical heterogeneity.
Grad-CAM scores $\Gamma^\mathrm{L}_\mathrm{CNN}$ extracted by CNN~(Supplementary Figure~\ref{fig:sup-snap2d}c) and Voronoi volumes $\Upsilon$~(Supplementary Figure~\ref{fig:sup-snap2d}d) both showed correlations with dynamics when they were coarse-grained~\cite{Liu_2024}.

\begin{figure}[h]
	\centering
	\includegraphics[width=0.5\textwidth]{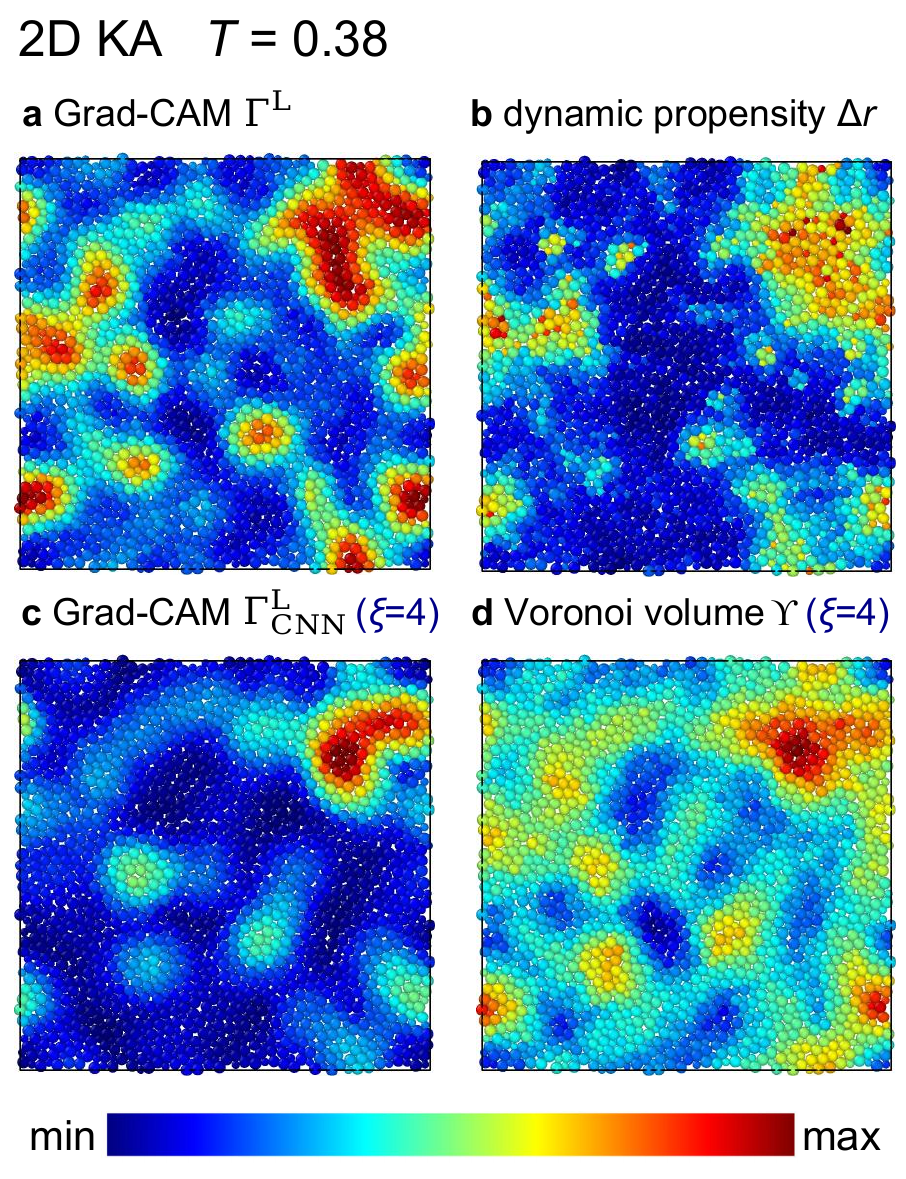} %
	\caption{\textbf{Snapshots of static and dynamical quantities in 2D KA mixture.}
    Snapshots of \textbf{a} Grad-CAM scores based on GNN $\Gamma^\mathrm{L}$, \textbf{b} dynamic propensity $\Delta r$, \textbf{c} Grad-CAM scores based on CNN $\Gamma^\mathrm{L}_\mathrm{CNN}$, and \textbf{d} Voronoi volume $\Upsilon$ at a low temperature of $T = 0.38$ in 2D KA model. \textbf{c} and \textbf{d} are coarse-grained with a coarse-graining length $\xi$ of 4.    
    All the values increase with the color scale, ranging from blue (minimum) to red (maximum).    
    Higher values of $\Gamma^\mathrm{L}$ indicate more liquid-like structures, whereas lower values indicate more glass-like structures. 
    Higher values of $\Delta r$ correspond to more mobile particles, whereas lower values correspond to more immobile particles.}
	\label{fig:sup-snap2d} %
\end{figure}

\subsection*{Scaled dynamics}
\begin{figure*}
	\centering
	\includegraphics[width=1\textwidth]{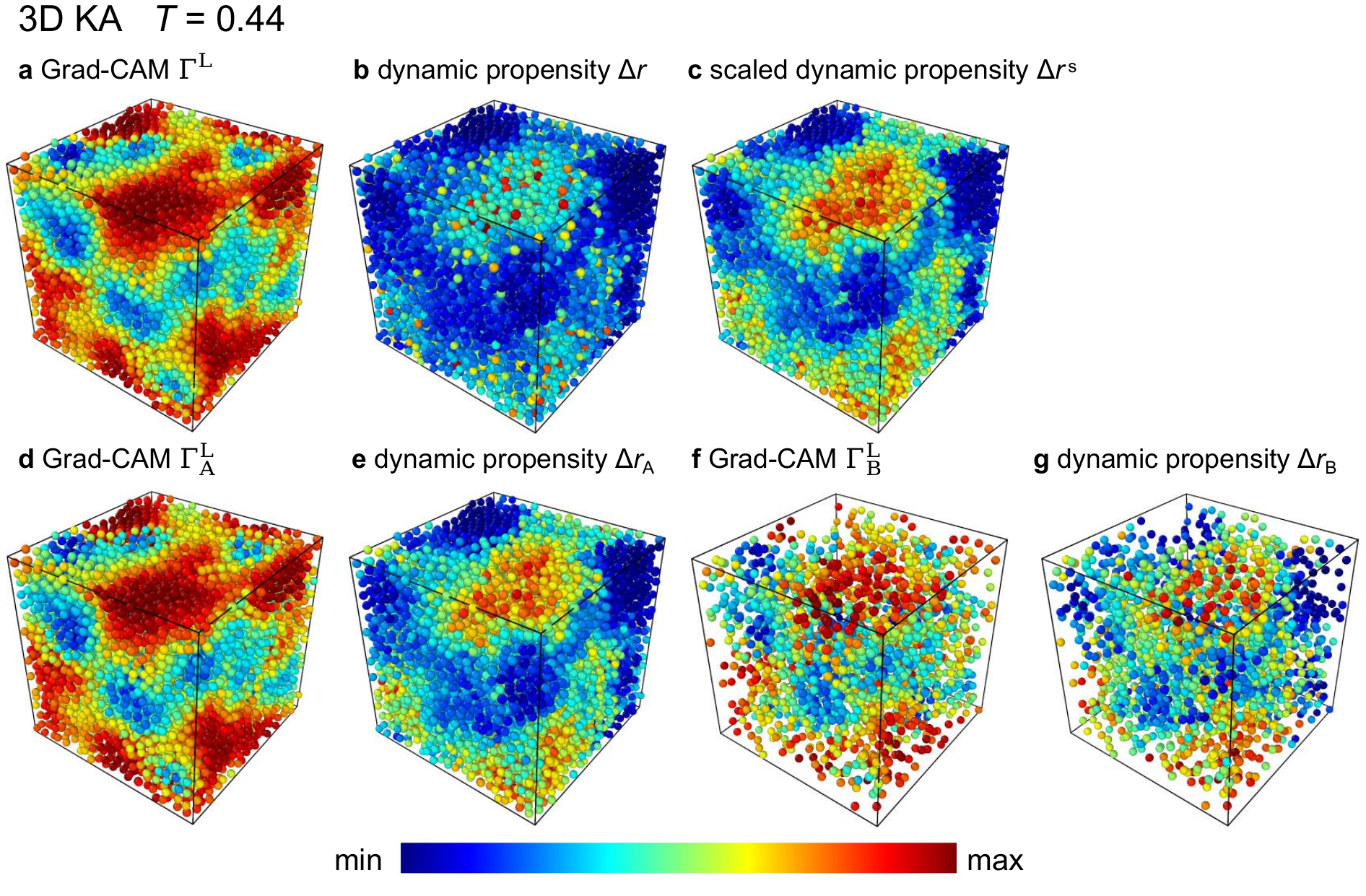}%
	\caption{\textbf{Snapshots of Grad-CAM scores and dynamic propensity for different species in 3D KA model.} 
    Above: snapshots of \textbf{a} Grad-CAM scores $\Gamma^\mathrm{L}$, \textbf{b} bare dynamic propensity $\Delta r$, \textbf{c} scaled dynamic propensity $\Delta r^\mathrm{s}$.
    Below: snapshots of \textbf{d} Grad-CAM scores for particles of type A, $\Gamma^\mathrm{L}_\mathrm{A}$, \textbf{e} bare dynamic propensity for particles of type A, $\Delta r_\mathrm{A}$, \textbf{f} Grad-CAM scores for particles of type B, $\Gamma^\mathrm{L}_\mathrm{B}$, and \textbf{g} bare dynamic propensity for particles of type B, $\Delta r_\mathrm{B}$.
    We used the same configuration at $T=0.44$ with $N=10000$ as that in Fig.~3 of the main text.
    All the quantities range from blue to red, as shown by the color bar.}
    \label{fig:sup-scale_dy} 
\end{figure*}

In 3D KA and 3D WAH systems, as shown in Supplementary Table~\ref{tab:cor}, the correlation coefficients $C_{\Gamma^\mathrm L,~\Delta r}$ between Grad-CAM scores $\Gamma^\mathrm{L}$ (see snapshot in Supplementary Figure~\ref{fig:sup-scale_dy}a) and bare dynamic propensity $\Delta r$ (snapshot in Supplementary Figure~\ref{fig:sup-scale_dy}b) calculated over all particles were relatively low.
However, when analyzing the two particle species separately, the correlations became more pronounced~\cite{Boattini_2020ML, Paret_2020ML}. 
Specifically, the correlations $C^\mathrm A_{\Gamma^\mathrm L,~\Delta r}$ between Grad-CAM scores $\Gamma^\mathrm{L}_\mathrm{A}$ (Supplementary Figure~\ref{fig:sup-scale_dy}d) and dynamic propensity $\Delta r_\mathrm{A}$ (Supplementary Figure~\ref{fig:sup-scale_dy}e) for A particles, and $C^\mathrm B_{\Gamma^\mathrm L,~\Delta r}$ between $\Gamma^\mathrm{L}_\mathrm{B}$ (Supplementary Figure~\ref{fig:sup-scale_dy}f) and $\Delta r_\mathrm B$ (Supplementary Figure~\ref{fig:sup-scale_dy}g) for B particles, were both higher than $C_{\Gamma^\mathrm L,~\Delta r}$.
Therefore, in Figs.~3d, h and 8c, f of the main text, we used the snapshots of scaled dynamic propensity $\Delta r^{\mathrm{s}}_i=\Delta r_i/ \langle \Delta r_{\alpha(i)} \rangle$, with $\langle \Delta r_{\alpha(i)} \rangle$ the average dynamic propensity of the species to which particle \(i\) belongs~(shown in Supplementary Figure~\ref{fig:sup-scale_dy}c).
In Figs.~5a, b and 8a, d of the main text, we calculated the species-weighted average correlations $C^{\mathrm{av}}_{\Gamma^\mathrm L,~\Delta r} = (N_\mathrm{A} C^\mathrm{A}_{\Gamma^\mathrm L,~\Delta r}+N_\mathrm{B} C^\mathrm{B}_{\Gamma^\mathrm L,~\Delta r})/N$.
And the values of $C^{\mathrm{av}}_{\Gamma^\mathrm L,~\Delta r}$ and  $C_{\Gamma^\mathrm L,~\Delta r_\mathrm{s}}$ were close to each other.
While in 3D WCA-HS model, $C_{\Gamma^\mathrm L,~\Delta r}$ was higher than both $C^\mathrm A_{\Gamma^\mathrm L,~\Delta r}$ and $C^\mathrm B_{\Gamma^\mathrm L,~\Delta r}$, so we directly used the snapshots of $\Delta r$ in Figs.~4d, h and ~8i of the main text, and plotted $C_{\Gamma^\mathrm L,~\Delta r}$ in Figs.~5c and 8g of the main text.

\begin{table}[htbp]
\centering
\caption{\textbf{Correlation coefficients between Grad-CAM scores and dynamic propensity.}
From left to right: data for $T=0.44$ in 3D KA, $T=0.56$ in 3D WAH, and $\eta^*=0.583$ in 3D WCA-HS systems.
Superscripts A and B denote the two particle species. $C^\mathrm{av}_{\Gamma^\mathrm{L},\Delta r}$ is the species-weighted average correlation coefficient, and $\Delta r^\mathrm{s}$ is the scaled dynamic propensity defined in the text. $t$ denotes the timescale for dynamic propensity $\Delta r$. }
\label{tab:cor}
\vspace{6pt}
\begin{tabular}{lcccccc}
\hline
system & \multicolumn{2}{c}{3D KA} & \multicolumn{2}{c}{3D WAH}& \multicolumn{2}{c}{3D WCA-HS} \\\hline
$t$&$\tau_\alpha$&$5\tau_\alpha$&$\tau_\alpha$&$8\tau_\alpha$&$\tau_\alpha$&$5\tau_\alpha$ \\\hline
$C_{\Gamma^\mathrm{L},\Delta r(t)}$ &0.259&0.154& 0.442 & 0.471 & 0.473 & 0.536\\
$C^\mathrm{A}_{\Gamma^\mathrm{L},\Delta r(t)}$ &0.448&0.589&0.512&0.588&0.378&0.473\\
$C^\mathrm{B}_{\Gamma^\mathrm{L},\Delta r(t)}$ &0.521&0.607&0.490&0.545&0.452&0.522\\
$C^\mathrm{av}_{\Gamma^\mathrm{L},\Delta r(t)}$ &0.463&0.592&0.501&0.566&0.430&0.507\\
$C_{\Gamma^\mathrm{L},\Delta r^\mathrm{s}(t)}$ &0.461&0.588&0.496&0.561&0.421&0.489\\
\hline
\end{tabular}
\end{table}

\begin{figure*}
	\centering
	\includegraphics[width=1\textwidth]{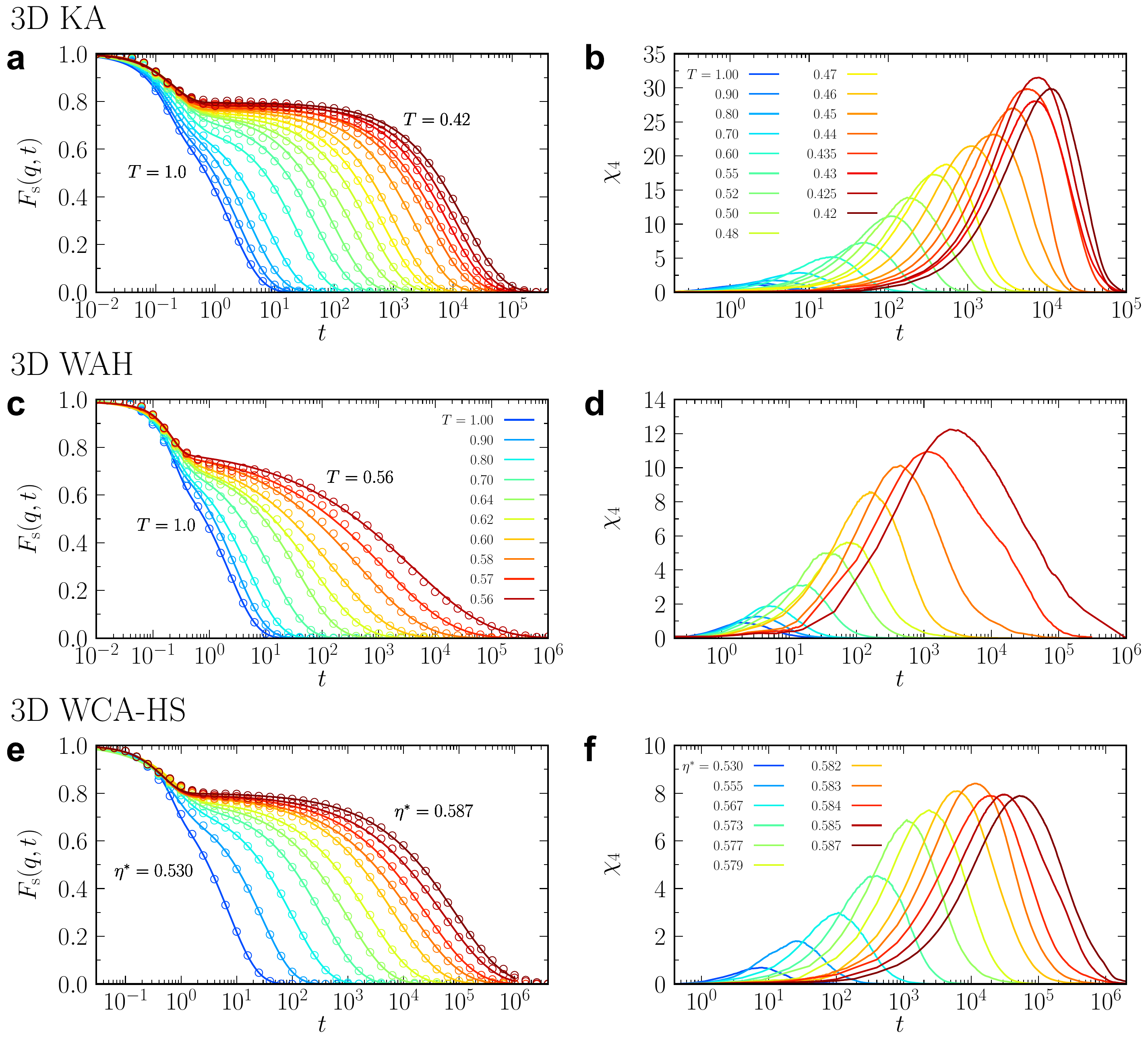}%
	\caption{\textbf{Dynamics of supercooled liquids for three systems.}
    \textbf{a} The self-intermediate scattering functions $F_\mathrm{s}(q,t)$ with $q=7.25$ for A particles as a function of time $t$ at different temperatures (indicated by colors) in the 3D KA model.
    The data~(symbols) were fitted with a double stretched exponential function~(curves).
    \textbf{b} The dynamic susceptibility $\chi_4(t)$ for A particles as a function of time $t$ at different temperatures in the 3D KA model.
    \textbf{c}, \textbf{d} The same as panels \textbf{a} and \textbf{b}, but in the 3D WAH system with $q=7.5$.
    \textbf{e}, \textbf{f} The same as \textbf{a} and \textbf{b}, but for different eﬀective packing fractions in the 3D WAC-HS system with $q=7.20-7.44$.
    }
    \label{fig:sup-dy} 
\end{figure*}

\begin{figure*}
	\centering
	\includegraphics[width=1\textwidth]{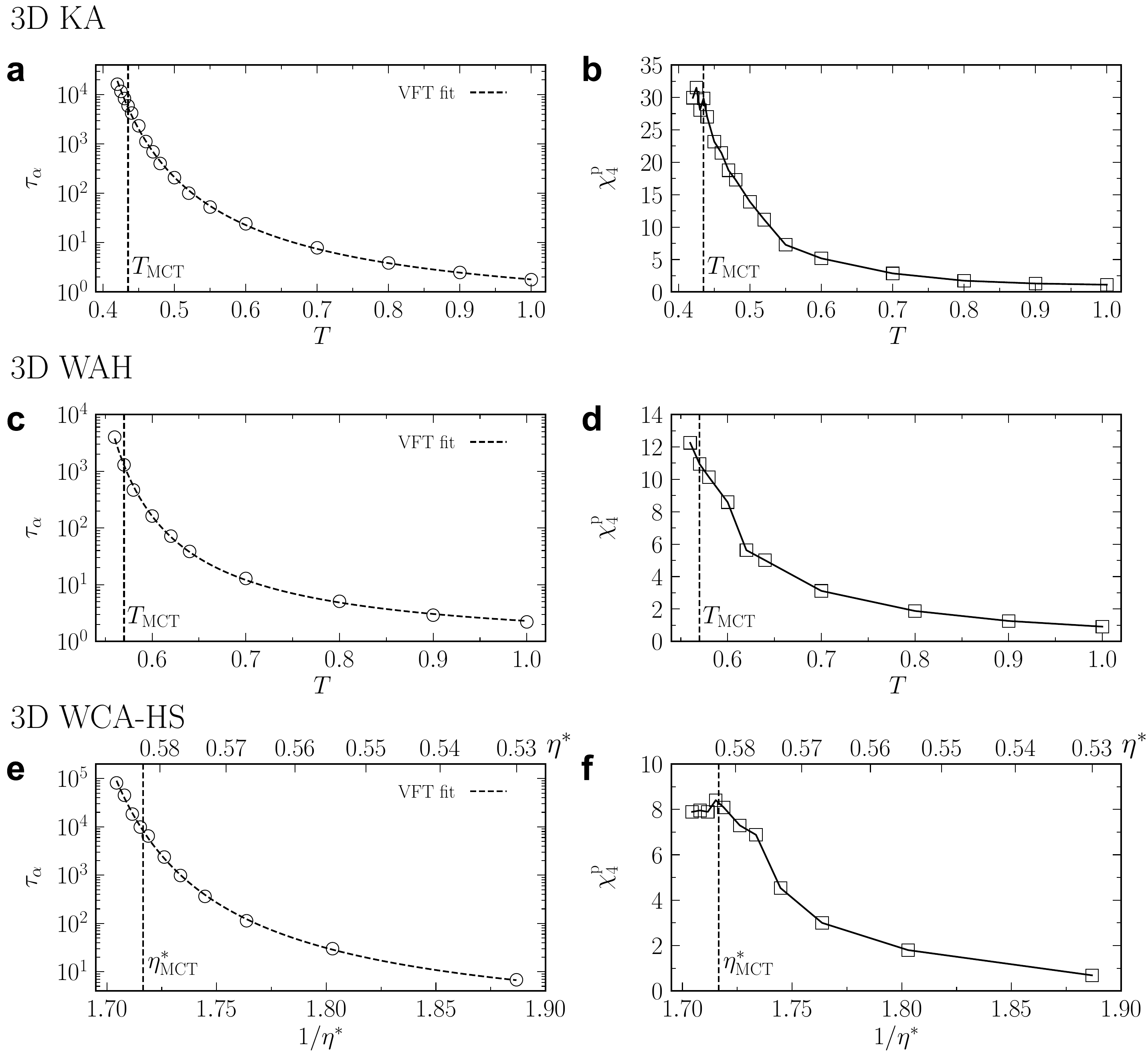}
	\caption{\textbf{Dynamical parameters of supercooled liquids for three systems.} 
    Temperature $T$~(or eﬀective packing fraction $\eta^*$) dependence of relaxation time $\tau_\alpha$~(open circles) obtained via fitting data in Supplementary Figure~\ref{fig:sup-dy}, for \textbf{a} 3D KA, \textbf{c} 3D WAH, and \textbf{e} 3D WAC-HS models. 
    We plotted a VFT fit on $\tau_\alpha$ with $\tau_\alpha= \tau_0\mathrm{exp}[\frac{1}{B(T/T_0-1)}]$ for \textbf{a}~($\tau_0=0.271$, $B=0.2283$ and $T_0=0.30$) and \textbf{c}~($\tau_0=0.726$, $B=0.8371$ and $T_0=0.49$), and with $\tau_\alpha= \tau_0\mathrm{exp}[\frac{1}{B(\eta^*_0/\eta^*-1)}]$~($\tau_0=0.515$, $B=2.790$ and $\eta^*_0=0.604$) for \textbf{e}.
    Temperature $T$~(or packing fraction $\eta^*$) dependence of $\chi_4^\mathrm{p}$, the peak values of the dynamic susceptibility, for \textbf{b} 3D KA, \textbf{d} 3D WAH, and \textbf{f} 3D WAC-HS systems.
    The values of eﬀective packing fraction $\eta^*$ in \textbf{e} and \textbf{f} are also shown at the top.
    The vertical dashed line indicates the mode-coupling temperature $T_{\mathrm{MCT}}$ (or mode-coupling packing fraction $\eta^*_{\mathrm{MCT}}$). }
    \label{fig:sup-dyp} 
\end{figure*}

\subsection*{Slow dynamics and dynamical heterogeneity}
To quantitatively study the slow dynamics of supercooled liquids, we calculated the self-intermediate scattering functions $F_\mathrm{s}(q,t)$ of A particles:
\begin{equation}\label{eq:fs}
\begin{aligned}
F_\mathrm{s}(q,t) &= \frac{1}{N_\mathrm{A}}\sum_{i=1}^{N_A}\langle e^{i\boldsymbol{q}\cdot(\boldsymbol{r}_i(t)-\boldsymbol{r}_i(0))}\rangle.
\end{aligned}
\end{equation}
Here, the wavenumber was set to the value at which the structure factor $S(q)$ reached its maximum value~\cite{Kob_1995sd, Lacevic_2003chi4}.
We plotted the self-intermediate scattering functions $F_\mathrm{s}(q,t)$ of A particles versus time $t$ at different temperatures~(or packing fractions) for systems of 3D KA in Supplementary Figure~\ref{fig:sup-dy}a, 3D WAH in Supplementary Figure~\ref{fig:sup-dy}c, and 3D WCA-HS in Supplementary Figure~\ref{fig:sup-dy}e.

At high temperatures or low packing fractions~(blue symbols on the left), $F_\mathrm{s}(q,t)$ quickly decayed to $0$.
As the system was cooled or compressed~(yellow and red symbols on the right), $F_\mathrm{s}(q,t)$ exhibited the two-step relaxation behavior with a plateau.
Our results support the drastic slowdown of dynamics in supercooled liquids reported in previous numerical studies~\cite{Angell_1995g,Kob_1995KAM,Kob_1995sd}.
We fitted the data to a double stretched exponential function, $F_\mathrm{s}(q,t)=f\exp(-(t/\tau_1)^{\beta_1})+(1-f)\exp(-(t/\tau_\alpha)^{\beta_2})$~($0\leq \beta_2 \leq1$)~\cite{Sengupta_2013chi4}, and extracted the relaxation time $\tau_\alpha$.
The temperature~(or packing fraction) dependence of $\tau_\alpha$~(in Supplementary Figure~\ref{fig:sup-dyp}a, c, and e) can be fitted by the Vogel–Fulcher–Tammann~(VFT) equation~\cite{Sengupta_2013chi4} yielding the ideal glass transition temperatures $T_0$ or packing fraction $\eta^*_0$~(see the specific data in the caption of Supplementary Figure~\ref{fig:sup-dyp}).
The fitted $T_0$ and $\eta^*_0$ agreed well with previous studies~\cite{Sengupta_2013chi4,Das_2022chi4}.

We also investigated dynamical heterogeneity in terms of the four-point dynamic susceptibility, $\chi_4(t)$, which has often been studied in previous works~\cite{Lacevic_2003chi4,Das_2022chi4}.
We defined $\chi_4(t)$ based on the fluctuations in overlap function $Q_4(t)$ for particles of type A as:
\begin{equation}\label{eq:chi4}
\chi_{4}(t) = \frac{1}{N_\mathrm{A}}\left[ \left< \left( Q_4(t) \right)^2 \right> - \left< Q_4(t) \right>^2 \right],
\end{equation}
where $N$ in $Q_4(t)$ defined in equation~(5) of the main text is replaced by $N_\mathrm{A}$ here.
Supplementary Figure~\ref{fig:sup-dy} (b, d, and f) displays the four-point dynamic susceptibility $\chi_4(t)$ of A particles for different systems.
$\chi_4(t)$ was first enhanced with time $t$ and reached a peak value around a specific time scale.
The peak value $\chi^\mathrm{p}_4$ characterizes the dynamical heterogeneity.
As illustrated in Supplementary Figure~\ref{fig:sup-dyp}b, d, and f, with decreasing temperatures~(or compression), $\chi^\mathrm{p}_4$ increased gradually at high temperatures or low packing fractions and then more rapidly in the deeply supercooled regime, indicating the growth of dynamical heterogeneity.
For 3D KA and 3D WCA-HS models in Supplementary Figure~\ref{fig:sup-dyp}b, f, however, $\chi^\mathrm{p}_4$ got saturated below $T_{\mathrm{MCT}}$ or above $\eta^*_{\mathrm{MCT}}$ , consistent with previous reports~\cite{ortlieb_2023,Das_2022chi4,Liu_2024}.

\subsection*{Message-passing-iteration number in GNN model}
\begin{figure}
    \centering
    \includegraphics[width=0.4\textwidth]{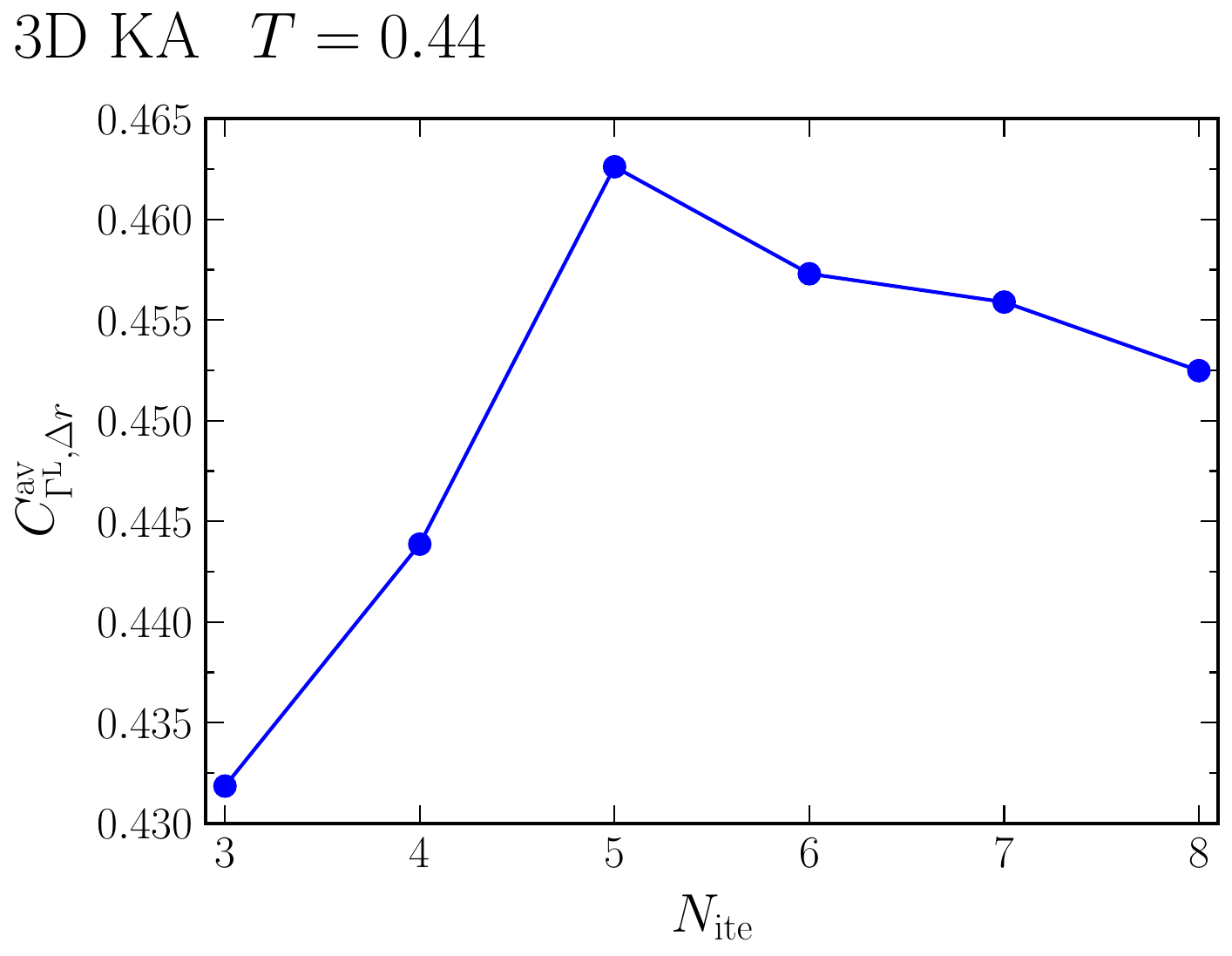}
    \caption{\textbf{$N_\mathrm{ite}$ dependence of GNN performance at \(T=0.44\) in 3D KA model.}
    Pearson's correlation coefficients $C^{\mathrm{av}}_{\Gamma^\mathrm L,~\Delta r}$ between the Grad-CAM scores \( \Gamma^{\mathrm L} \) and the dynamic propensity \(\Delta r\) as a function of message-passing-iteration-number $N_\mathrm{ite}$ for edge and node updates in GNN training.
    }
    \label{fig:sup-nite}
\end{figure}

To construct the GNN model, we specified the number of message-passing iterations $N_\mathrm{ite}$ for updating edge and node features (see Fig.~1 in the main text). 
We found that classification accuracy remains 50\%~(unable to classify) when $N_\mathrm{ite} \leq 2$.
We then tested several values of $N_\mathrm{ite}$ larger than 2 and found that the correlation coefficient $C^{\mathrm{av}}_{\Gamma^\mathrm L,~\Delta r}$ was the highest at \(N_\mathrm{ite}=5\) in our study. 
Supplementary Figure~\ref{fig:sup-nite} shows an example for the 3D KA mixture at $T=0.44$.
The correlation coefficient changed only slightly, from 0.462 to 0.452, as \(N_\mathrm{ite}\) increased from 5 to 8, indicating that the results are relatively insensitive to the choice of \(N_\mathrm{ite}\) within this range.

\subsection*{Analysis of the inherent structure}
\begin{figure}
	\centering
	\includegraphics[width=0.5\textwidth]{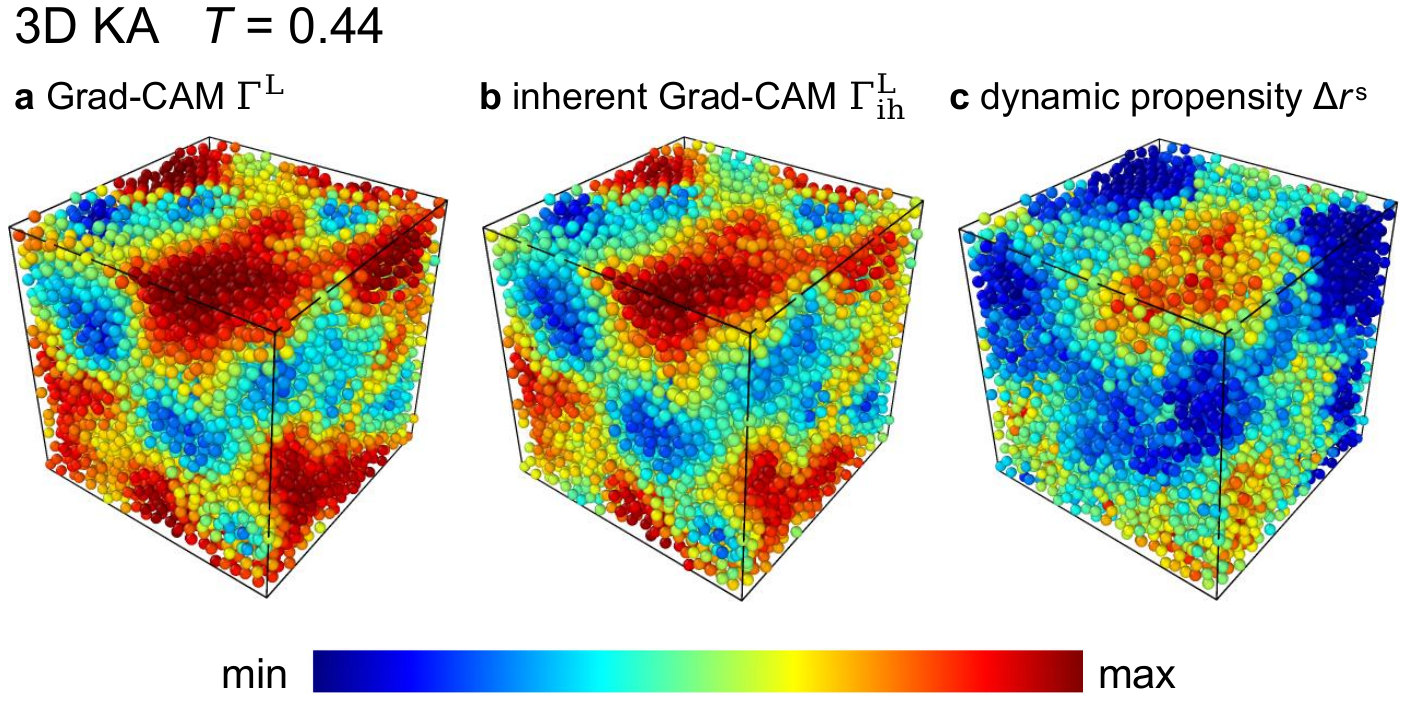}
	\caption{\textbf{Comparisons to Grad-CAM scores for inherent structures at $T=0.44$ in 3D KA model.}
    \textbf{a} Snapshot of Grad-CAM scores $\Gamma^\mathrm{L}$ for the instantaneous configuration.
    \textbf{b} Snapshot of Grad-CAM scores $\Gamma_\mathrm{ih}^\mathrm{L}$ for the inherent structure of the configuration in \textbf{a}.
    \textbf{c} Snapshot of scaled dynamic propensity $\Delta r^\mathrm{s}$.
    All parameters increase from blue to red as indicated by the color bar.
    }
	\label{fig:sup-ih} 
\end{figure}

Previous works~\cite{Schoenholz_2016ML, Jung_2023, Liu_2024} found that dynamics correlated with the inherent structure underlying the equilibrium~(instantaneous) supercooled-liquid state, so we also analyzed the inherent state.
We mention that only the configurations for structural characterization were replaced with their inherent structures, while the dynamic propensity and analysis protocol remained the same as in the main study.
The Grad-CAM scores for inherent states (Supplementary Figure~\ref{fig:sup-ih}b) displayed a similar spatial distribution to equilibrium Grad-CAM scores (Supplementary Figure~\ref{fig:sup-ih}a) and dynamic propensity (Supplementary Figure~\ref{fig:sup-ih}c).
According to the data shown in Supplementary Table~\ref{tab:cor-ih}, the inherent Grad-CAM scores $\Gamma^\mathrm{L}$ exhibited stronger correlations with dynamic propensity $\Delta r$ compared to the instantaneous ones.
Correlations between coarse-grained potential energy $U$ and $\Delta r$ were also enhanced when using inherent structure information.
Regarding potential-trained Grad-CAM scores $\Gamma_U^\mathrm{L}$ and coarse-grained Voronoi volume $\Upsilon$, the inherent and equilibrium configurations showed nearly equal correlation coefficients with dynamics. 
These results reveal that our method can be applied to the inherent structure to extract the structural features that correlate with dynamics.

\begin{table}[htbp]
\centering
\caption{\textbf{Structure--dynamics correlation coefficients for instantaneous and inherent structures at $T=0.44$ in 3D KA mixture.} 
From top to bottom: correlation coefficients between dynamic propensity $\Delta r$ and Grad-CAM scores $\Gamma^\mathrm L$, Grad-CAM scores $\Gamma_U^\mathrm L$ obtained by training potential, coarse-grained Voronoi volume $\Upsilon(\xi=3)$, and coarse-grained potential energy $U(\xi=3)$.
Columns 2 and 3 correspond to correlations for instantaneous configurations, and columns 4 and 5 correspond to those for inherent structures. 
$t$ indicates the timescale for $\Delta r$.}
\label{tab:cor-ih}
\vspace{6pt}
\begin{tabular}{lcccc}
\hline
state & \multicolumn{2}{c}{instantaneous} & \multicolumn{2}{c}{inherent}\\\hline
$t$&$\tau_\alpha$&$5\tau_\alpha$&$\tau_\alpha$&$5\tau_\alpha$\\\hline
$C^\mathrm{av}_{\Gamma^\mathrm{L},\Delta r(t)}$ &0.463&0.592&0.503&0.628\\
$C^\mathrm{av}_{\Gamma_U^\mathrm{L},\Delta r(t)}$ &0.541&0.650&0.535&0.637\\
$C^\mathrm{av}_{\Upsilon(\xi=3),\Delta r(t)}$ &0.422&0.542&0.422&0.543\\
$C^\mathrm{av}_{U(\xi=3),\Delta r(t)}$ &0.519&0.582&0.571&0.646\\
\hline
\end{tabular}
\end{table}

\subsection*{Extending GNNs under different training conditions to systems of varying sizes}
\begin{figure*}
	\centering
	\includegraphics[width=1\textwidth]{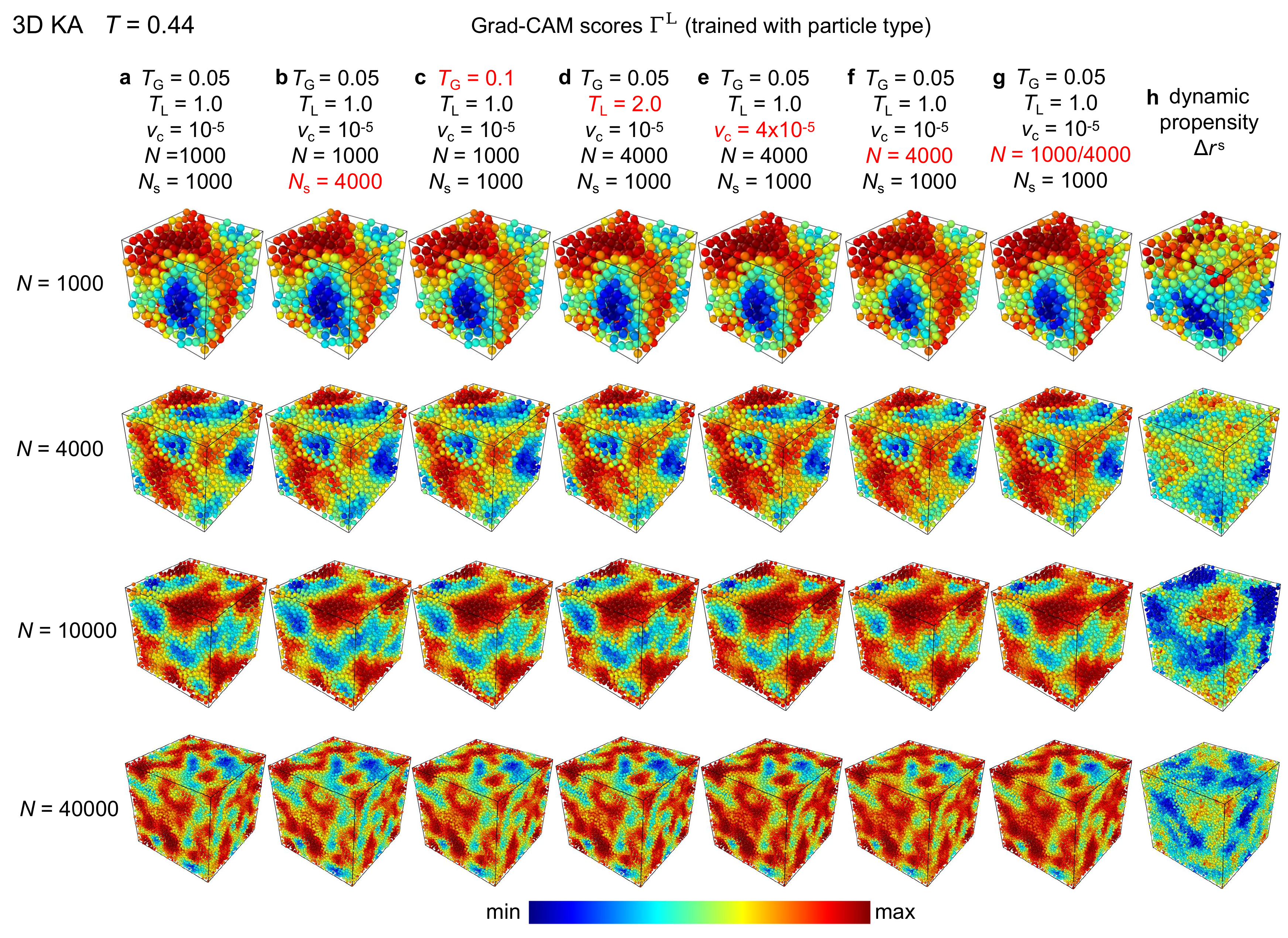}
	\caption{\textbf{Application of GNN trained on particle type under different training data conditions to systems of varying sizes at $T=0.44$ in 3D KA model.}
    Snapshots of Grad-CAM scores $\Gamma^\mathrm{L}$ obtained by training using only particle type as a node feature on different datasets:
    \textbf{a} the training dataset discussed in the main text as reference~(glass preparation temperature $T_\mathrm{G}=0.05$, liquid preparation temperature $T_\mathrm{L}=1.0$, cooling rate $v_\mathrm{c}=10^{-5}$, particle number $N=1000$, and training sample number $N_\mathrm{s}=1000$),
    \textbf{b} $N_\mathrm{s}$ increased to 4000, \textbf{c} $T_\mathrm{G}$ changed to 0.1, \textbf{d} $T_\mathrm{L}$ increased to 2.0, \textbf{e} $v_\mathrm{c}$ changed to $4\times10^{-5}$, \textbf{f} $N$ increased to 4000, and \textbf{g} $N$ equally combined of 1000 and 4000.
    The modified parameters are marked in red.
    \textbf{h} Snapshots of scaled dynamic propensity $\Delta r^\mathrm{s}$.
    The rows correspond to test samples with increasing particle numbers: $N=$1000, 4000, 10000, and 40000 from top to bottom.
    All the values increase from blue to red as indicated by the color bar.}
	\label{fig:sup-extend} 
\end{figure*}

\begin{figure*}
	\centering
	\includegraphics[width=1\textwidth]{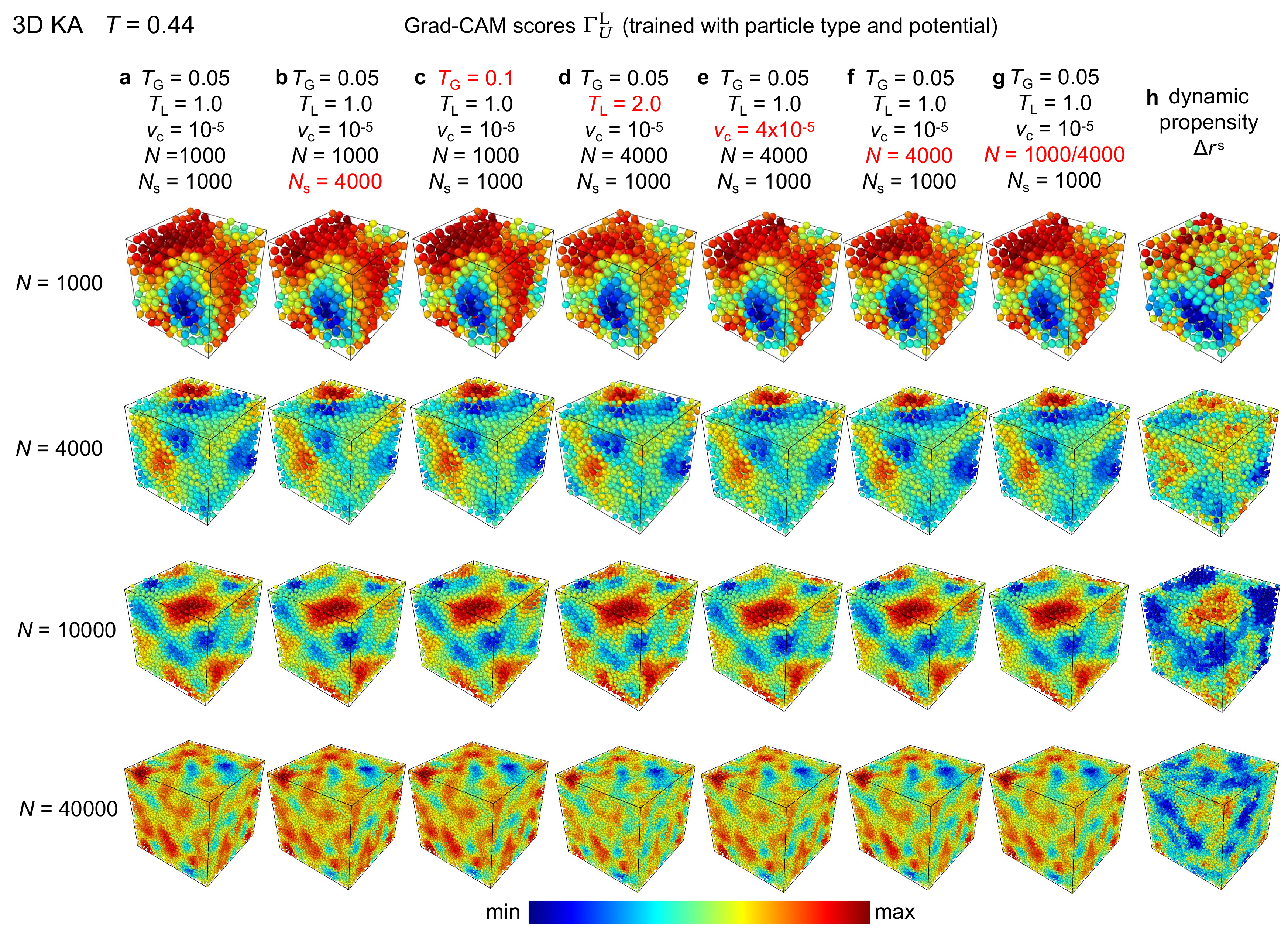}
	\caption{\textbf{Application of GNN trained on particle type and potential energy under different training data conditions to systems of varying sizes at $T=0.44$ in 3D KA model.}
    The same as Supplementary Figure~\ref{fig:sup-extend}, but Grad-CAM scores $\Gamma^\mathrm{L}_U$ were obtained by training with particle type and an additional indicator: the local potential energy \(U\).
    }
	\label{fig:sup-extendpot} 
\end{figure*}

Supplementary Figure~\ref{fig:sup-extend}a shows snapshots of Grad-CAM scores $\Gamma^\mathrm{L}$ of samples at $T=0.44$ in the 3D KA model with increasing particle numbers $N=1000$, 4000, 10000, and 40000 from top to bottom.
The GNN was trained on the same datasets: $N_\mathrm{s}=1000$ samples for liquid and glass~(each consisting of $N=1000$ particles), respectively, in which liquid was equilibrated at $T=1.0$, and glass was obtained at temperature $T_\mathrm{G}=0.05$ with cooling rate $v_\mathrm{c}=10^{-5}$.
As we can see, the structural heterogeneity characterized by $\Gamma^\mathrm{L}$ was evident across all samples.
When compared to the dynamic propensity $\Delta r^\mathrm{s}$ of each sample~(in Supplementary Figure~\ref{fig:sup-extend}h), we can observe correlations between $\Gamma^\mathrm{L}$ and $\Delta r^\mathrm{s}$.
The red parts of $\Gamma^\mathrm{L}$~(liquid-like) corresponded to red regions of $\Delta r^\mathrm{s}$~(fast dynamics) for all the samples.
So our GNN and Grad-CAM method can work widely for systems of different sizes (including much larger ones) to extract local structural features and predict dynamics, even though the GNN was trained only on small configurations. 

Additionally, we changed the GNN training datasets to check their influence on the results.
One training data condition was modified to: sample number $N_\mathrm{s}=4000$ (Supplementary Figure~\ref{fig:sup-extend}b), glass preparation temperature $T_\mathrm{G}=0.1$ (Supplementary Figure~\ref{fig:sup-extend}c), liquid preparation temperature $T_\mathrm{L}=2.0$ (Supplementary Figure~\ref{fig:sup-extend}d), cooling rate $v_\mathrm{c}=4\times10^{-5}$ (Supplementary Figure~\ref{fig:sup-extend}e), particle number $N=4000$ (Supplementary Figure~\ref{fig:sup-extend}f) and a equal combination of both $N=1000$ and $N=4000$ (Supplementary Figure~\ref{fig:sup-extend}g), respectively, while all other parameters remained unchanged~(as indicated by the labels above the snapshots).
In each column, we present the corresponding snapshots of $\Gamma^\mathrm{L}$ for the same test sample as those in Supplementary Figure~\ref{fig:sup-extend}a~(particle numbers increase from top to bottom).
Notably, all the snapshots of $\Gamma^\mathrm{L}$ under different training configurations for the same samples~(in the same row) were essentially identical.
In other words, our Grad-CAM results are remarkably robust against the training conditions.

Moreover, the Grad-CAM results based on GNN trained with both particle type and potential are also shown in Supplementary Figure~\ref{fig:sup-extendpot}, following the arrangement in Supplementary Figure~\ref{fig:sup-extend}.
The columns correspond to training datasets of Grad-CAM scores $\Gamma^\mathrm{L}$ or dynamic propensity $\Delta r^\mathrm{s}$, and the rows correspond to the particle numbers of test samples.
From these snapshots, we can again confirm that our method can work for samples of distinct sizes and is insensitive to the conditions of learning data, as explained previously.
With the inclusion of potential energy during GNN learning, Grad-CAM scores seemed to demonstrate clearer correlations with dynamic propensity than those in Supplementary Figure~\ref{fig:sup-extend}~(see Fig.~8 of the main text).

\subsection*{Radial correlation functions of glass-like structures}
\begin{figure*}
    \centering
    \includegraphics[width=\textwidth]{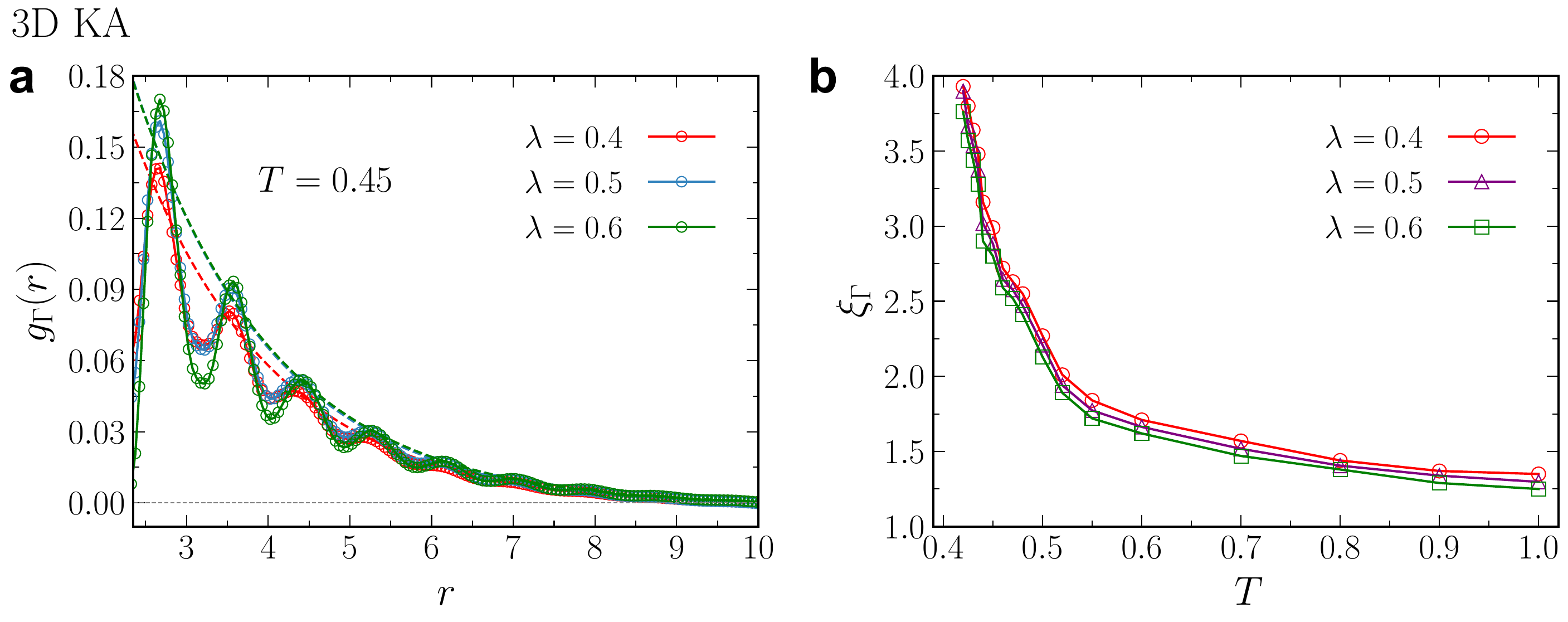}
    \caption{\textbf{Radial correlation functions and correlation lengths of glass-like structures with different thresholds in 3D KA model.}
    \textbf{a} Radial correlation functions $g_{\Gamma}(r)$ of glass-like structures identified by Grad-CAM scores for $T=0.45$ in 3D KA model. Different colors represent $g_{\Gamma}(r)$ obtained using different thresholds, $\Gamma^\mathrm{L}_c = \Gamma^\mathrm{L}_\mathrm{min} + \lambda\,(\Gamma^\mathrm{L}_\mathrm{max} - \Gamma^\mathrm{L}_\mathrm{min})$, with different $\lambda$. 
    The dashed curves are the exponential ``envelope fit" proportional to $e^{(-r/\xi_\Gamma)}$, with $\xi_\Gamma$ the correlation length. 
    \textbf{b} The fitted correlation lengths $\xi_\Gamma$ as functions of temperature $T$ for different values of $\lambda$.
    }
    \label{fig:sup-xigamma}
\end{figure*}

We defined the glass-like structures by the threshold $\Gamma^\mathrm{L}_c = \Gamma^\mathrm{L}_\mathrm{min} + \lambda\,(\Gamma^\mathrm{L}_\mathrm{max} - \Gamma^\mathrm{L}_\mathrm{min})$ with $\lambda = 0.5$~(see equation~(9) in Methods of the main text).
Here, we also considered $\lambda = 0.4$ and $0.6$, and calculated the corresponding radial correlation functions $g_{\Gamma}(r)$ to compare with those obtained using $\lambda = 0.5$ at $T=0.45$ in the 3D KA mixture (Supplementary Figure~\ref{fig:sup-xigamma}a).
More particles were regarded as being in the glass-like state when $\Gamma^\mathrm{L}_c$ was larger~(with a larger value of $\lambda$).
$g_{\Gamma}(r)$ with $\lambda = 0.6$ exhibited relatively higher peaks, whereas $g_{\Gamma}(r)$ with $\lambda = 0.4$ showed lower peaks than those with $\lambda = 0.5$. 
The correlation length scales $\xi_\Gamma$ can be obtained by fitting $g_{\Gamma}(r)$, and they were very similar for all three values of $\lambda$.
In all cases, $\xi_\Gamma$ increased with the decrease of temperatures (Supplementary Figure~\ref{fig:sup-xigamma}b).

We further checked $g_{\Gamma}(r)$ scaled by radial distribution functions $g(r)=\frac{1}{N\rho}\sum_{l \neq k}^{N}\left\langle\delta\!\left(\mathbf{r}_l-\mathbf{r}_k-\mathbf{r}\right)\right\rangle$, which was define as:
\begin{equation}\label{eqs:g_gamma}
\begin{split}
	g^\mathrm{s}_{\Gamma}(r)={}&\frac{1}{g(r)N\rho}\sum_{l \neq k}^{N}
	\left\langle \omega_l^\Gamma \omega_k^\Gamma
	\delta\!\left(\mathbf{r}_l-\mathbf{r}_k-\mathbf{r}\right)\right\rangle 
	-\left\langle \frac{1}{N}Q_\Gamma \right\rangle^2 ,
\end{split}
\end{equation}
where $\omega_i^\Gamma$ was determined using $\lambda = 0.5$.
The scaled $g^\mathrm{s}_{\Gamma}(r)$ and corresponding fitted correlation lengths $\xi^\mathrm{s}_\Gamma$ for the 3D KA system are shown in Supplementary Figure~\ref{fig:sup-xigamma-gr}.
$\xi^\mathrm{s}_\Gamma$ based on scaled $g^\mathrm{s}_{\Gamma}(r)$ increased with decreasing temperatures, which behaved qualitatively the same as $\xi_\Gamma$ based on $g_{\Gamma}(r)$.

\begin{figure*}
    \centering
    \includegraphics[width=\textwidth]{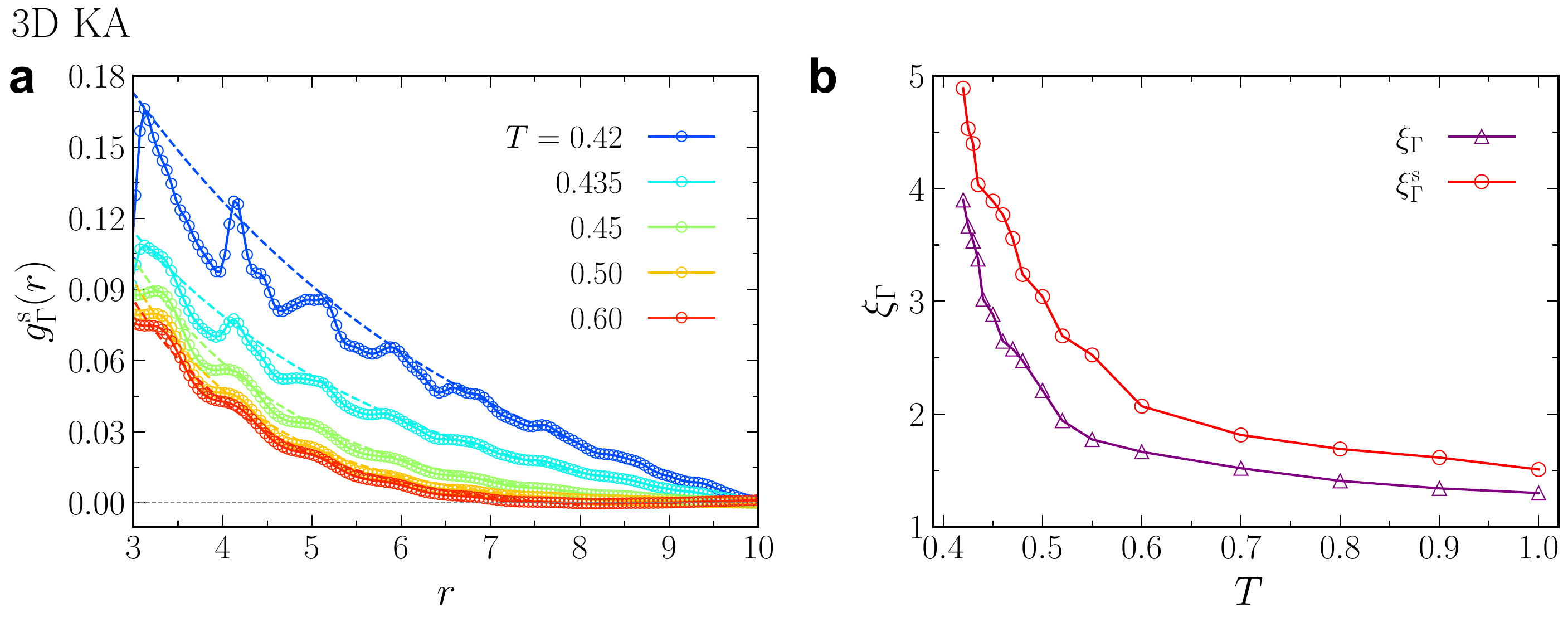}
    \caption{\textbf{Scaled radial correlation functions and correlation lengths of glass-like structures in 3D KA model.}
    \textbf{a} Scaled radial correlation functions $g^\mathrm{s}_{\Gamma}(r)$ of glass-like structures at different temperatures in the 3D KA model, with exponential ``envelope fits" (dashed curves).
    \textbf{b} The correlation lengths $\xi^\mathrm{s}_\Gamma$ obtained by fitting scaled $g^\mathrm{s}_{\Gamma}(r)$ on the left~(red symbols) and $\xi_\Gamma$ obtained by fitting $g_{\Gamma}(r)$~(purple symbols; see Fig.~7a of the main text) against temperatures $T$.
    }
    \label{fig:sup-xigamma-gr}
\end{figure*}

\subsection*{Further study on structure--dynamics correlations}
As shown in Supplementary Figure~\ref{fig:sup-cor}, we studied coarse-grained potential energy \( U\) and coarse-grained Voronoi volume \(\Upsilon\) with different coarse-graining lengths \(\xi\). 
Their correlations with dynamics reached the maximum values at specific lengths~($\xi=3$ for 3D KA and 3D WAH and $\xi=2$ for 3D WCA-HS systems), which were used for the correlation coefficients in Fig.~8 of the main text.
Additionally, correlations between structures and dynamic propensity \(\Delta r\) for longer timescales were investigated.
They continued to increase after $\tau_\alpha$ and then started to saturate or decrease at some timescales, $t_\mathrm{max}$, which was similar to the data reported in previous papers~\cite{Bapst_2020GNN,Boattini_2021ML}.
We put the correlations $C^{\mathrm{av}}_{\Gamma^\mathrm L,~\Delta r}$ for $\Delta r(t)$ over $t=t_\mathrm{max}$~($t_\mathrm{max}=5\tau_\alpha$ for 3D KA and 3D WCA-HS and $t_\mathrm{max}=8\tau_\alpha$ for 3D WAH models) in Figs.~5 and 8 of the main text for reference~(see Supplementary Tables~\ref{tab:cor} and~\ref{tab:cor-ih} for specific coefficients).

\begin{figure*}
    \centering
    \includegraphics[width=\textwidth]{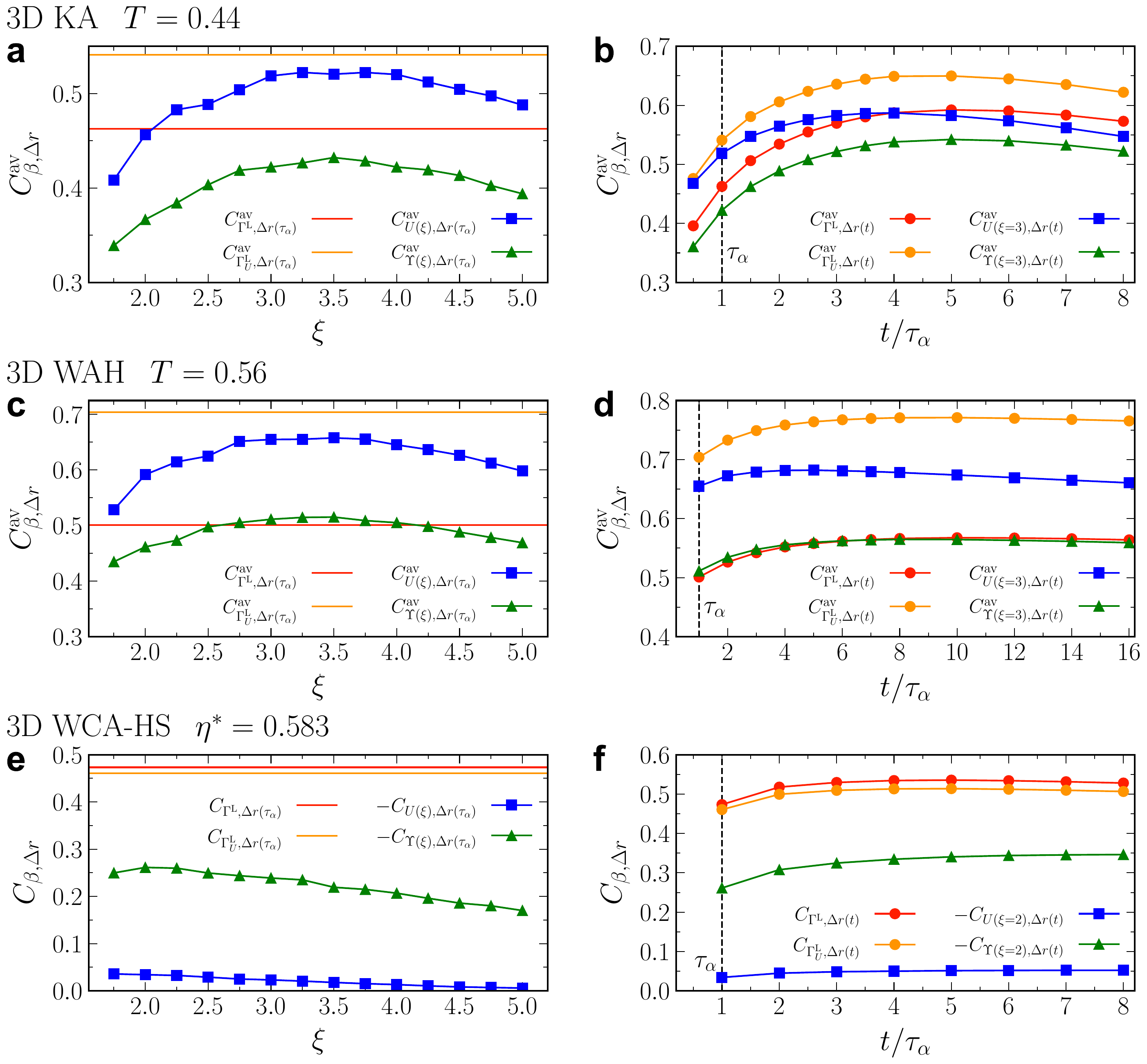}
    \caption{\textbf{Correlations between structural and dynamical fields for three systems.}
    Data for systems of \textbf{a}, \textbf{b} 3D KA at \(T=0.44\), \textbf{c}, \textbf{d} 3D WAH at \(T=0.56\), and \textbf{e}, \textbf{f} 3D WCA-HS at \(\eta^*=0.583\).
    \textbf{a}, \textbf{c}, and \textbf{e} Pearson's correlation coefficients between dynamic propensity \(\Delta r\) over \(\tau_\alpha\) and coarse-grained potential energy \( U\), as well as coarse-grained Voronoi volume \(\Upsilon\), as functions of the coarse-graining lengths \(\xi\). 
    The correlation coefficients between dynamic propensity \(\Delta r\) and Grad-CAM scores \( \Gamma^{\mathrm L} \) (red line) and the Grad-CAM scores obtained by training potential \( \Gamma_U^{\mathrm L} \) (orange line) are also shown (both without coarse-graining).
    \textbf{b}, \textbf{d}, and \textbf{f} Correlation coefficients between structural indicators and dynamic propensity \(\Delta r\) over time $t$, plotted as functions of timescale $t/\tau_\alpha$.
    The vertical dashed line marks the timescale of $t=\tau_\alpha$.
    $U$ and $\Upsilon$ were coarse-grained with a fixed \(\xi\) (\(\xi=3\) for panels \textbf{b} and \textbf{d} and \(\xi=2\) for panel \textbf{f}), around which they showed almost the highest correlations with dynamics.}
    \label{fig:sup-cor}
\end{figure*}

\end{document}